\documentclass[preprint,longabstract]{aastex}
\usepackage{lineno}
\usepackage{times}
\usepackage{amsmath}
\usepackage{color}

\begin{document}

\slugcomment{ApJ}

\title{Searches for Large-Scale Anisotropy in the Arrival Directions of Cosmic Rays \\
Detected above Energy of $10^{19}$\,eV \\
at the Pierre Auger Observatory and the Telescope Array}

\author{
A.~Aab\altaffilmark{42a}, 
P.~Abreu\altaffilmark{65a}, 
M.~Aglietta\altaffilmark{54a}, 
E.J.~Ahn\altaffilmark{83a}, 
I.~Al Samarai\altaffilmark{29a}, 
I.F.M.~Albuquerque\altaffilmark{17a}, 
I.~Allekotte\altaffilmark{1a}, 
J.~Allen\altaffilmark{87a}, 
P.~Allison\altaffilmark{89a}, 
A.~Almela\altaffilmark{11a,8a}
% ...
}
\affil{The Pierre Auger Collaboration}
% ...
\altaffiltext{1a}{Centro At\'{o}mico Bariloche and Instituto Balseiro (CNEA-UNCuyo-CONICET), San Carlos de Bariloche, Argentina.}
% ...
\altaffiltext{8a}{Instituto de Tecnolog\'{\i}as en Detecci\'{o}n y Astropart\'{\i}culas (CNEA, CONICET, UNSAM), Buenos Aires, Argentina.}
% ...
\altaffiltext{11a}{Universidad Tecnol\'{o}gica Nacional - Facultad Regional Buenos Aires, Buenos Aires, Argentina.}
% ...
\altaffiltext{17a}{Universidade de S\~{a}o Paulo, Instituto de F\'{\i}sica, S\~{a}o Paulo, SP, Brazil.}
% ...
\altaffiltext{29a}{Institut de Physique Nucl\'{e}aire d'Orsay (IPNO), Universit\'{e} Paris 11, CNRS-IN2P3, Orsay, France.}
% ...
\altaffiltext{42a}{Universit\"{a}t Siegen, Siegen, Germany.}
% ...
\altaffiltext{54a}{Osservatorio Astrofisico di Torino  (INAF), Universit\`{a} di Torino and Sezione INFN, Torino, Italy.}
% ...
\altaffiltext{65a}{Laborat\'{o}rio de Instrumenta\c{c}\~{a}o e F\'{\i}sica Experimental de Part\'{\i}culas - LIP and  Instituto Superior T\'{e}cnico - IST, Universidade de Lisboa - UL, Portugal.}
% ...
\altaffiltext{83a}{Fermilab, Batavia, IL, USA.}
% ...
\altaffiltext{87a}{New York University, New York, NY, USA.}
% ...
\altaffiltext{89a}{Ohio State University, Columbus, OH, USA.}
% ...

\and

\author{
R.U.~Abbasi\altaffilmark{1t},
M.~Abe\altaffilmark{13t},
T.Abu-Zayyad\altaffilmark{1t},
M.~Allen\altaffilmark{1t},
R.~Anderson\altaffilmark{1t},
R.~Azuma\altaffilmark{2t},
E.~Barcikowski\altaffilmark{1t},
J.W.~Belz\altaffilmark{1t},
D.R.~Bergman\altaffilmark{1t},
S.A.~Blake\altaffilmark{1t}
% ...
}
\affil{The Telescope Array Collaboration}
\altaffiltext{1t}{High Energy Astrophysics Institute and Department of Physics and Astronomy, University of Utah, Salt Lake City, Utah, USA.}
\altaffiltext{2t}{Graduate School of Science and Engineering, Tokyo Institute of Technology, Meguro, Tokyo, Japan.}
% ...
\altaffiltext{8t}{Institute for Cosmic Ray Research, University of Tokyo, Kashiwa, Chiba, Japan.}
% ...
\altaffiltext{11t}{Faculty of Engineering, Kanagawa University, Yokohama, Kanagawa, Japan.}
% ...
\altaffiltext{13t}{The Graduate School of Science and Engineering, Saitama University, Saitama, Saitama, Japan.}
% ...
\altaffiltext{17t}{Institute for Nuclear Research of the Russian Academy of Sciences, Moscow, Russia.}
% ...
\altaffiltext{29t}{National Institute of Radiological Science, Chiba, Chiba, Japan.}
% ...

\begin{abstract}
Spherical harmonic moments are well-suited for capturing anisotropy at any scale in the flux of cosmic rays. 
An unambiguous measurement of the full set of spherical harmonic coefficients requires full-sky coverage.
This can be achieved by combining data from observatories located in both the northern and southern 
hemispheres. To this end, a joint analysis using data recorded at the Telescope Array and the Pierre Auger 
Observatory above $10^{19}$~eV is presented in this work. The resulting multipolar expansion of the 
flux of cosmic rays allows us to perform a series of anisotropy searches, and in 
particular to report on the angular power spectrum of cosmic rays above $10^{19}$~eV. No significant
deviation from isotropic expectations is found throughout the analyses performed. Upper limits on the 
amplitudes of the dipole and quadrupole moments are derived as a function of the direction in the sky,
varying between 7\% and 13\% for the dipole and between 7\% and 10\% for a symmetric
quadrupole.
\end{abstract}



\section{Introduction}
\label{sec:intro}

The large-scale distribution of arrival directions of cosmic rays is an important observable in attempts
to understand their origin. This is because this observable is closely connected to both their source 
distribution and propagation. Due to scattering in magnetic fields, the anisotropy imprinted 
upon the distribution of arrival directions is mainly expected at large scales up to the highest energies. 
Large-scale patterns with anisotropy contrast at the level of $10^{-4}$ to $10^{-3}$ have been reported by 
several experiments for energies below $10^{15}$\,eV where the high flux of cosmic rays allows 
the collection of a large number of events~\citep{tibet,SK,eastop,milagro,icecubea,icecubeb,icecubec,icecubed}. 
For energies above a few $10^{15}$\,eV, the decrease of the flux with energy makes it challenging 
to collect the necessary statistics required to reveal amplitudes down to $10^{-3}$ or $10^{-2}$.
Upper limits at the level of a few percents have been obtained at 
$\simeq 10^{16}$~eV~\citep{kascadegrande} and $\simeq 10^{18}$~eV~\citep{auger-ls,auger-ls3d}.

The anisotropy of any angular distribution on the sphere is encoded in the corresponding set of 
spherical harmonic moments $a_{\ell m}$. Although not predictable in a quantitative way at present, 
large-scale anisotropies might be expected from various mechanisms of propagation of cosmic rays. 
A non-zero dipole moment is naturally expected from propagation models leading to a cosmic ray 
density gradient embedding the observer. Even in the absence of a density gradient, a measurable dipole 
moment might result from the motion of the Earth or of the massive objects in the neighborhood 
of the Milky-Way relative to a possibly stationary cosmic-ray rest frame~\citep{compton,kachelriess,harari}.
On the other hand, excesses along a plane, for instance the super-Galactic one, would be detectable 
as a prominent quadrupole. The dipole and the quadrupole moments are thus of special interest, but 
an access to the full set of multipoles is relevant to characterize departures from isotropy at all scales. 
However, since cosmic ray observatories at ground level have only a partial-sky coverage, the 
recovering of these multipoles turns out to be nearly impossible without explicit assumptions on the 
shape of the underlying angular distribution~\citep{sommers}. Indeed, for an angular distribution 
described by a multipolar expansion bounded to some degree $\ell$, the multipole coefficients can only 
be estimated within a resolution degraded exponentially with $\ell$~\citep{billoir}. In most cases, given the 
available statistics, only the dipole and the quadrupole moments can be estimated with a relevant 
resolution under the assumption that the flux of cosmic rays is purely dipolar or purely dipolar and 
quadrupolar, respectively. Evading such hypotheses and thus measuring the multipoles to any order 
in an unambiguous way requires \emph{full-sky coverage}. At present, full-sky coverage can only be 
achieved through the meta-analysis of data recorded at observatories located in both hemispheres.

The Telescope Array and the Pierre Auger Observatory are the two largest experiments ever built to study
ultra-high energy cosmic rays in each hemisphere. The aim of the joint analysis reported in this 
article is to search for anisotropy with full-sky coverage by combining the data of the two experiments. The data 
sets used for this search are described in section~\ref{sec:observatories} together with some properties 
and performances of the experiments relevant for this study. Special emphasis is given to the control 
of the event counting rate with time and local angles as well as to the respective exposures to each direction 
of the sky. To facilitate this first joint analysis, the energy threshold used in this report, 
$10^{19}$\,eV, is chosen to guarantee that both observatories operate with full detection efficiency for any 
of the events selected in each data set. Above this energy, the respective exposure functions follow purely 
geometric expectations. 

The analysis method to estimate the spherical harmonic moments $a_{\ell m}$ is presented in 
section~\ref{sec:a_lm}, together with its statistical performance. The main challenge in combining the data 
sets is to account adequately for the relative exposures of both experiments. The empirical approach adopted 
here is shown in section~\ref{sec:jointmethod} to meet the challenge. 
Results of the estimated multipole coefficients are then presented and illustrated in several ways in 
section~\ref{sec:dataanalysis}, with, in particular, reports for the first time of a significance full-sky map of 
the overdensities and underdensities and of the angular power spectrum above $10^{19}$\,eV. Several 
cross-checks against systematic effects are presented in section~\ref{sec:systematics}, showing the 
robustness of the analyses. Finally, the results are discussed in section~\ref{sec:conclusions}, together 
with some prospects for future joint analyses.

\section{The Observatories and the Data Set}
\label{sec:observatories}

\subsection{The Pierre Auger Observatory}
\label{sec:auger}

The Pierre Auger Observatory, from which data taking started in 2004 and which has been fully operational since 
January 2008, is located in the Southern hemisphere in Malarg\"{u}e, Argentina (mean latitude 
$-35.2^\circ$)~\citep{auger-exp}. It consists of 1660 water-Cherenkov detectors laid out over about 
3000\,km$^2$ overlooked by 27 fluorescence telescopes grouped in five buildings. The 
hybrid nature of the Pierre Auger Observatory enables the assignment of the energy of each event to be 
derived in a calorimetric way through the calibration of the shower size measured with the surface detector 
array by the energy measured with the fluorescence telescopes on a subset of high quality hybrid 
events~\citep{auger-hybrid}. 

The data set used in this study consists of events recorded by the surface detector array from 1 January 2004 up 
to 31 December 2012 with zenith angles up to $60^\circ$. Optimal angular and energy reconstructions are ensured 
by requiring that all six neighbors of the water-Cherenkov detector with the highest signal were active at the time each 
event was recorded~\citep{auger-exposure}. Based on this condition, the angular resolution is of about 
${\simeq}1^\circ$~\citep{bonifazi}; while the energy resolution above $10^{19}$\,eV amounts about to 
10\%~\citep{auger-hybrid} with a systematic uncertainty on the absolute energy scale of 14\%~\citep{verzi}. 
The full efficiency of the surface detector array is reached above $3{\times}10^{18}$\,eV~\citep{auger-exposure}. 
With a corresponding total exposure of 31\,440\,km$^2$\,sr\,yr, the total number of events above $10^{19}$\,eV is 8259.

\subsection{The Telescope Array}
\label{sec:ta}

The Telescope Array, which has been fully operational since March 2008, is located in the Northern 
hemisphere in Utah, USA (mean latitude $+39.3^\circ$). It consists of 507 scintillator detectors covering 
an area of approximately 700\,km$^2$~\citep{TA-SD} overlooked on dark nights by 38 fluorescence 
telescopes located at three sites~\citep{TA-FDa,TA-FDb}. The scintillator detector array allows the 
detection of cosmic rays with high duty cycle by sampling at the ground level the lateral distribution 
of the showers induced in the atmosphere. On the other hand, the fluorescence detectors are used 
to sample the longitudinal profiles of the showers, allowing a calorimetric measurement of the energy, as
with the Auger Observatory. 
The subset of events detected simultaneously by both detection techniques is then used to rescale
the energy of the events recorded by the scintillator detector array to the calorimetric estimate provided 
by the fluorescence detectors~\citep{TA-hybrid}.

The data set provided for the present study by the Telescope Array consists of events recorded between
11 May 2008 and 3 May 2013 with zenith angles smaller than $55^\circ$. The selection of the events is based 
on both fiducial and quality criteria. Each event must include at least five scintillator detectors (counters), and 
the counter with the largest signal must be surrounded by four working counters that are its nearest neighbors, 
excluding diagonal separation, on a 1200\,m grid. Both the timing 
and the lateral distribution fits of the signals must have $\chi^2/\text{ndf}$ value less than 4.
The angular uncertainty estimated by the timing fit must be less than $5^\circ$, and the fractional 
uncertainty in the shower size estimated by the lateral distribution fit must be less than 25\%.
Based on these criteria, the energy above which the surface scintillator array operates with full efficiency 
is ${\simeq}8{\times}10^{18}$\,eV. The energy resolution is better than 20\% above $10^{19}$\,eV with a systematic 
uncertainty on the absolute energy scale of 21\%~\citep{TA-escale}. The total exposure is 6040\,km$^2$\,sr\,yr, 
for a total number of events above $10^{19}$\,eV amounting to 2130.

\subsection{Control of the Event Rate}
\label{sec:rate}


The control of the event rate is of critical importance. The magnitude of the spurious pattern imprinted in the 
arrival directions by any effect of experimental origin must be kept under control. This is essential if this
magnitude is larger than the fluctuations on the anisotropy parameters intrinsic to the available statistics. 

The instantaneous exposure of each experiment is not constant in time due to 
the construction phase of the observatories and to
unavoidable dead times of detectors (\textit{e.g.} failures of electronics, power supply, 
communication system, ...). This translates into modulations of the event rates even for an isotropic flux.
However, these dead times concern only a few detectors and are randomly distributed over operation time,
so that once averaged over several years of data taking, the relative modulations of both exposure functions 
in \emph{local sidereal time} turn out to be not larger than a few per thousand.
The impact for anisotropy searches is thus expected to be negligible given the small statistics available.

\begin{figure}[!t]
  \centering					 
 \includegraphics[width=0.48\textwidth]{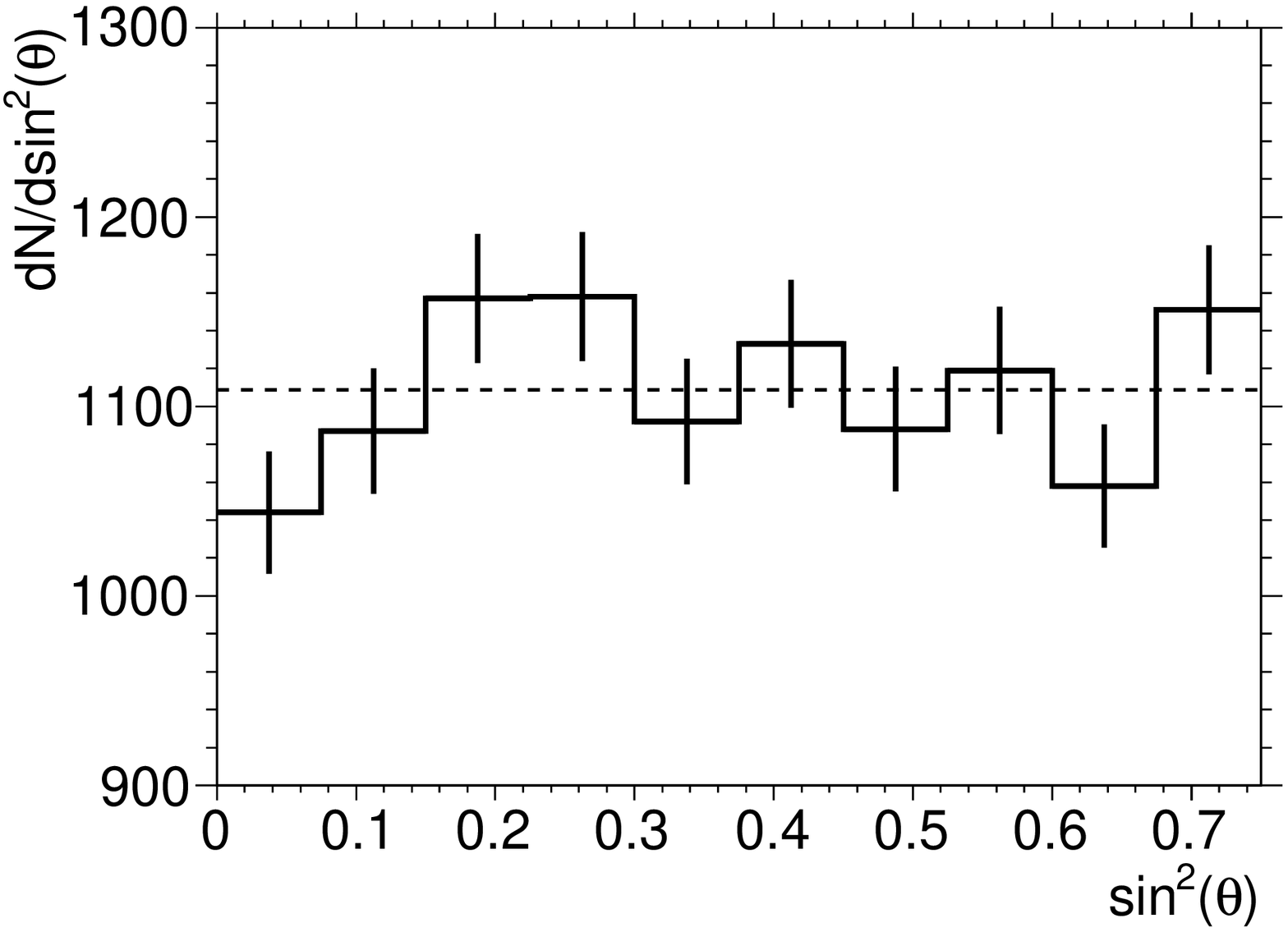}
 \includegraphics[width=0.48\textwidth]{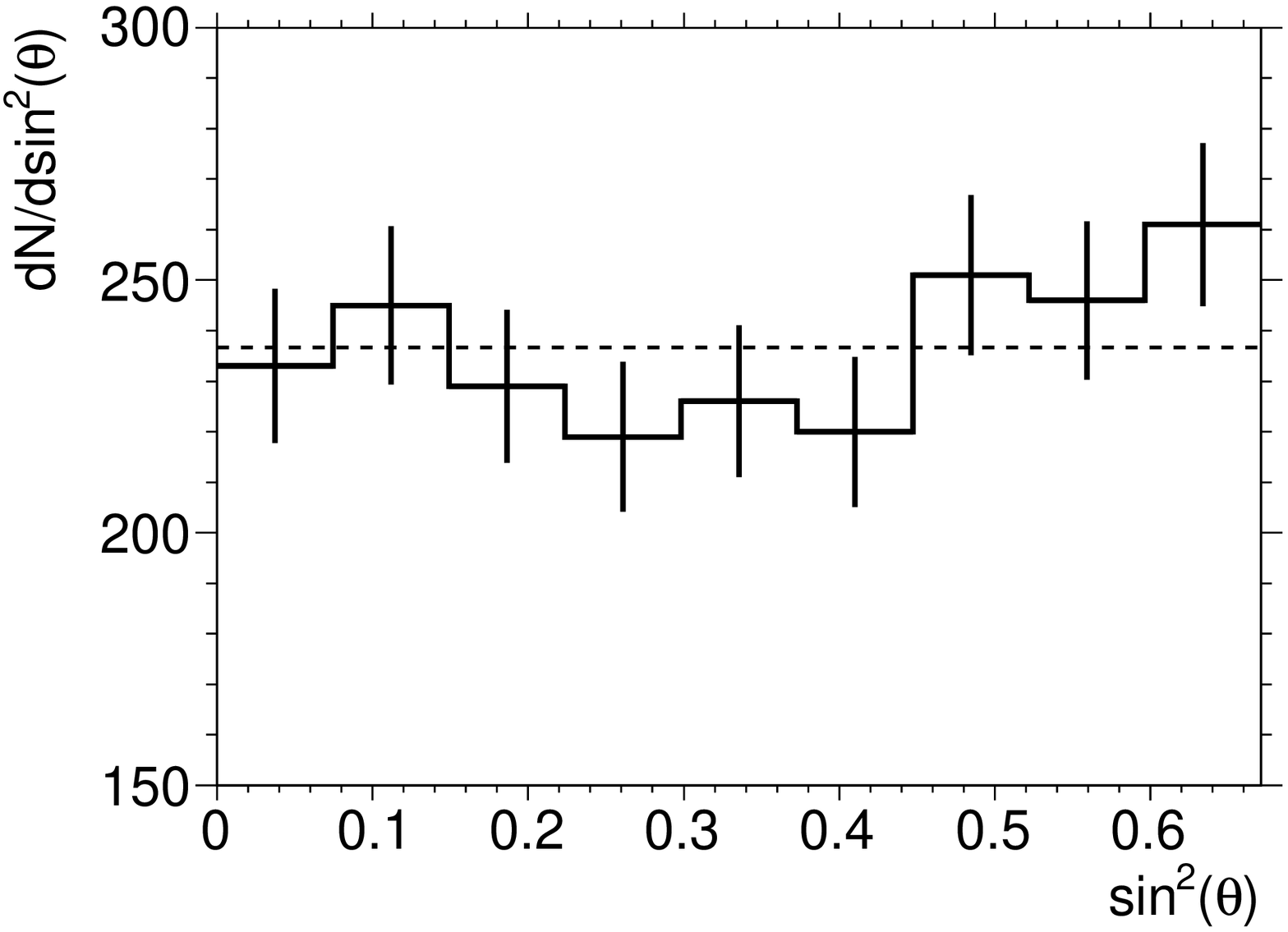}
  \caption{Distributions $\mathrm{d}N/\mathrm{d}\sin^2\theta$ above $10^{19}$\,eV for the data of the Pierre Auger Observatory (left) 
  and of the Telescope Array (right).The expected intensity levels are shown as the dotted lines.}
  \label{fig:sin2}
\end{figure}

In terms of local angles, the event counting rate is controlled by the relationship between the measured shower size 
and zenith angle used to estimate the energy by accounting for the attenuation of the showers in the atmosphere. 
For an isotropic flux and full efficiency, the distribution in zenith angles $\mathrm{d}N/\mathrm{d}\theta$ of a surface detector array is 
expected to be proportional to $\sin\theta\,\cos\theta$ for solid angle and geometry reasons, so that the distribution 
in $\mathrm{d}N/\mathrm{d}\sin^2\theta$ is expected to be uniform. Note that this distribution is quasi-invariant to large-scale 
anisotropies~\citep{auger-ls3d}, so that requiring the distributions $\mathrm{d}N/\mathrm{d}\sin^2\theta$ to be uniform constitutes 
a well-suited tool to control the event counting rate. Both distributions are shown in figure~\ref{fig:sin2}
to be indeed uniform above $10^{19}$\,eV within statistical uncertainties.

Due to the steepness of the energy spectrum, the event counting rate can also be largely distorted by systematic
changes of the energy estimate with time and/or local angles. The two main effects acting in this way, namely the 
atmospheric and geomagnetic effects, are by default accounted for in large-scale anisotropy searches reported by 
the Auger collaboration~\citep{auger-weather,auger-gmf}. This is necessary given the available statistics around 
$10^{18}$\,eV. However, above $10^{19}$\,eV, the impact of these effects is marginal given the reduced statistics. 
The fact that the same treatment is not implemented in the data set provided by the Telescope Array has no impact 
on the accuracy of the anisotropy measurements presented in this report, as is shown below.

Changes of atmospheric conditions are known to modulate the event rate as a function of time. This is because 
the development of an extensive air shower is sensitive to the atmospheric pressure and air density in a way which
influences the measurement of the shower size at a fixed distance from the core, and consequently the measurement
of the energy. To avoid the undesired variations of the event rate induced by these effects, the observed shower
size has to be related to the one that would have been measured at some fixed reference values of pressure and 
density. Such a procedure is implemented to produce the data set recorded at the Pierre Auger 
Observatory~\citep{auger-weather}.

Assuming there is no variation of the cosmic-ray flux over a time scale of few years,
the fact that the time windows of the data sets provided by the two observatories are different 
has no impact on the anisotropy searches presented in this report, given that more than three years of data are
considered in each set separately. With at least three years of data taking indeed, the event rate variation at 
the solar time scale \textit{decouples} from the one at the sidereal time scale, so that any significant 
modulation of the event rate of experimental origin, primarily visible at the solar time scale,
would not impact significantly the event rate at the sidereal time scale~\citep{letessier-selvon}.

\begin{figure}[!t]
  \centering					 
 \includegraphics[width=0.7\textwidth]{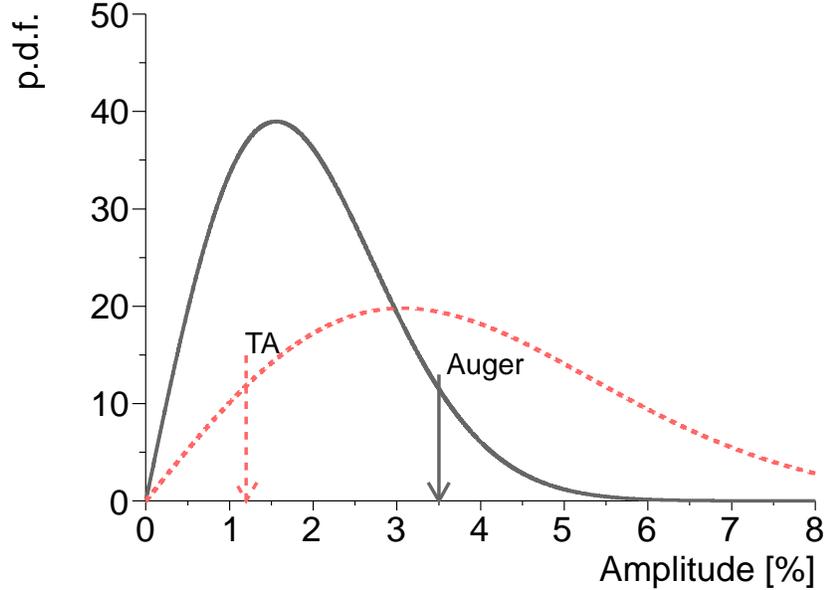}
 \caption{Amplitude of first harmonic at the anti-sidereal time scale measured with Auger (black solid line) and Telescope 
 Array (red dashed line) data. The curves are the background expectations from the respective Rayleigh distributions.}
 \label{fig:antisid}
\end{figure}

Although there is no shower size correction for weather effects in the data set of the Telescope Array, it is worth 
noting, however, that the natural time scale at which the modulation of the event rate operates is the \emph{solar} 
one. This means that over whole years of data taking as in the present analysis, such a modulation is expected to 
be partially compensated at the \emph{sidereal} time scale, as just emphasized. The size of the residual effect, 
together with the level of 
the sideband effect induced by a seasonal modulation of the daily counting rate, can be checked empirically by 
evaluating the amplitude of the first harmonics for fictitious right ascension angles calculated by dilating the time of 
the events in a way that a solar day lasts about four minutes longer~\citep{farley}. The corresponding time scale is 
called the \emph{anti-sidereal} one. A first harmonic amplitude standing out from the background noise at such an 
unphysical time scale would be indicative of important spurious effects of instrumental origin in the measurement 
of the first harmonic coefficients at the sidereal time scale. The measured values are shown in figure~\ref{fig:antisid} 
for each experiment, together with the respective Rayleigh distributions expected from statistical fluctuations. The 
amplitudes are seen to be compatible with that expected from the Rayleigh distributions. This provides support that 
the counting rate is not affected by spurious modulations of instrumental origin at the Pierre Auger Observatory - as 
expected from the corrections of the signal sizes -- \emph{and} at the Telescope Array as well.

Moreover, since we aim at characterizing the arrival directions in both right ascension and declination angles, it is
also important to control the event rate in terms of local angles. The geomagnetic field turns out to influence  
shower developments and shower size estimations at a fixed energy due to the broadening of the spatial
distribution of particles in the direction of the Lorentz force. The strength of the resulting modulation of the event rate
depends on the angle between the incoming direction of each event and the direction of the transverse component
of the geomagnetic field. The event rate is thus distorted independently of time as a function of local zenith and 
azimuth angles, and thus as a function of the declination. To eliminate these distortions, the data set recorded at the 
Pierre Auger Observatory is produced by relating the shower size of each event to the one that would have been 
observed in the absence of the geomagnetic field~\citep{auger-gmf}. However, the impact of these corrections is 
shown in section~\ref{sec:systematics} to be marginal 
given the limited statistics above $10^{19}$\,eV. That the same kind of corrections are not carried out on the data
set provided by the Telescope Array is thus unimportant for the present study. This is further reinforced by the fact
that geomagnetic effects are expected to be more important for a water-Cherenkov detector array which is more 
sensitive to muonic signal than for a scintillator one as measurements are made up to larger zenith angles.

\subsection{Directional Exposures}
\label{sec:exposures}

The directional exposure $\omega(\mathbf{n})$ provides the effective time-integrated collecting area for a flux from 
each direction of the sky. The small variations of the exposure in sidereal time 
translate into small variations of the directional exposure in right ascension. However, given the small size of these 
variations, the relative modulations of the respective directional exposure functions in right ascension turn out to be 
less than few $10^{-3}$. Given that the limited statistics currently available above $10^{19}$\,eV cannot allow 
an estimation of each $a_{\ell m}$ coefficient with a precision better than few percent, the non-uniformities of both 
$\omega_\text{TA}$ and $\omega_\text{Auger}$ in right ascension can be neglected. Both functions are 
consequently considered to depend only on the declination hereafter. 

\begin{figure}[!t]
  \centering					 
 \includegraphics[width=0.7\textwidth]{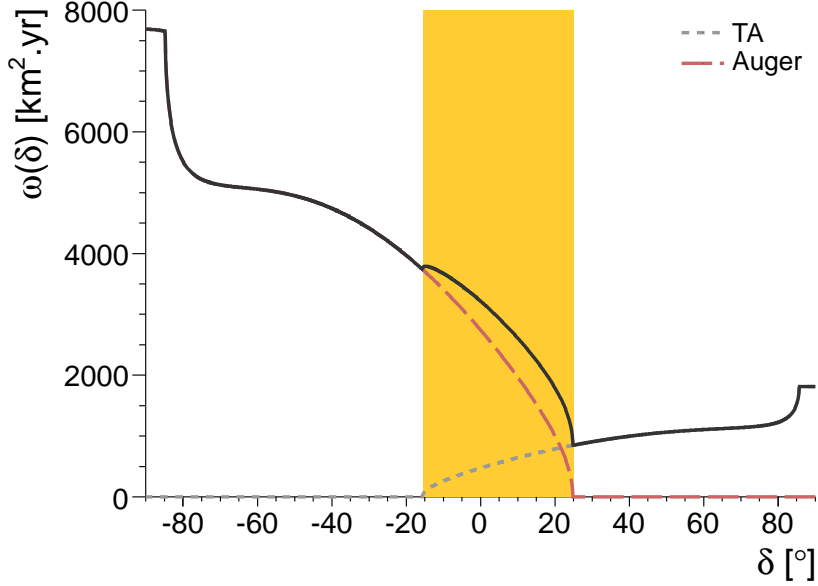}
  \caption{Directional exposure above $10^{19}$\,eV as obtained by summing the nominal 
  individual ones of the Telescope Array and the Pierre Auger Observatory, as a function of the declination. The 
  overlapping sky region is indicated by the yellow band.}
  \label{fig:exposure}
\end{figure}

Since the energy threshold of $10^{19}$\,eV guarantees that both experiments are fully efficient in their respective 
zenith range $[0,\theta_\text{max}]$, the directional exposure relies only on geometrical acceptance terms. 
The dependence on declination can then be obtained in an analytical way~\citep{sommers} as
\begin{equation}
\label{eqn:omegageom}
\omega_i(\mathbf{n})=A_i(\cos\lambda_i\,\cos\delta\,\sin\alpha_m+\alpha_m\sin\lambda_i\,\sin\delta),
\end{equation}
where $\lambda_i$ is the latitude of the considered experiment, the parameter $\alpha_m$ is given by
\begin{equation}
\label{eqn:alpham}
\alpha_m=\begin{cases}
0 & ;\xi>1,\\
\pi & ;\xi<-1,\\
\arccos\xi & ;\text{otherwise,}
\end{cases}
\end{equation}
with $\xi\equiv(\cos\theta_\text{max}-\sin\lambda_i\,\sin\delta)/\cos\lambda_i\,\cos\delta$,
and the normalization factors $A_i$ chosen such that the integration of each $\omega_i$ function over
$4\pi$ matches the (total) exposure of the corresponding experiment. The directional exposure functions 
$\omega_i(\delta)$ of each experiment are shown in figure~\ref{fig:exposure}. Given the respective latitudes 
of both observatories and with the maximum zenith angle used here, overall, it is clearly seen that full-sky 
coverage is indeed achieved when summing both functions. Also, and it will be important in the following, it 
is interesting to note that a common band of declination, namely $-15^\circ\leq\delta\leq 25^\circ$, is covered 
by both experiments.

In principle, the combined directional exposure of the two experiments should be simply the sum of the 
individual ones. However, individual exposures have here to be re-weighted by some empirical factor $b$ 
due to the unavoidable uncertainty in the relative exposures of the experiments:
\begin{equation}
\label{eqn:omega}
\omega(\mathbf{n};b)=\omega_\text{TA}(\mathbf{n})+b\,\omega_\text{Auger}(\mathbf{n}).
\end{equation}
Written in this way, $b$ is a dimensionless parameter of order unity. In practice, only an estimation 
$\bar{b}$ of the factor $b$ can be obtained, so that only an estimation of the directional exposure 
$\bar{\omega}(\mathbf{n})\equiv\omega(\mathbf{n};\bar{b})$ can be achieved through 
equation~\eqref{eqn:omega}. The procedure used for obtaining $\bar{b}$ from the joint data set is 
described in section~\ref{sec:jointmethod}. In addition, although the 
techniques for assigning energies to events are nearly the same, there are differences as to how the primary 
energies are derived at the Telescope Array and the Pierre Auger Observatory. Currently, systematic uncertainties 
in the energy scale of both experiments amount to about 21\% and 14\% respectively~\citep{TA-escale,verzi}. 
This encompasses the adopted fluorescence yield, the uncertainties for the absolute calibration of the 
fluorescence telescopes, the influence of the atmosphere transmission used in the reconstruction, the uncertainties 
in the shower reconstruction, and the uncertainties in the correction factor for the missing energy. Uncovering 
and understanding the sources of systematic uncertainties in the relative energy scale is beyond the scope of this 
report and will be addressed elsewhere. However, such a potential shift in energy leads to different counting rates 
above some fixed energy threshold, which induces fake anisotropies. Formally, these fake anisotropies are similar 
to the ones resulting from a shift in the relative exposures of the experiments\footnote{Note however that this 
statement is not exactly rigorous in the case of energy-dependent anisotropies in the underlying flux of cosmic rays, 
or in the case when the relative energy scale is zenith angle dependent.}. 
The parameter $b$ can thus be viewed as an effective correction which absorbs any kind of systematic uncertainties in the 
relative exposures, whatever the sources of these uncertainties. This empirical factor is arbitrarily chosen to re-weight 
the directional exposure of the Pierre Auger Observatory relatively to the one of the Telescope Array.

\section{Estimation of Spherical Harmonic Coefficients}
\label{sec:a_lm}

The observed angular distribution of cosmic rays, $\mathrm{d}N/\mathrm{d}\Omega$, can be naturally modeled as the sum 
of Dirac functions on the surface of the unit sphere whose arguments are the arrival directions 
$\{\mathbf{n}_1,\ldots,\mathbf{n}_N\}$ of the events,
\begin{equation}
\label{eqn:obsflux}
\frac{\mathrm{d}N(\mathbf{n})}{\mathrm{d}\Omega}=\sum_{i=1}^N\delta(\mathbf{n},\mathbf{n}_i).
\end{equation}
Throughout this section, arrival directions are expressed in the equatorial coordinate system (declination
$\delta$ and right ascension $\alpha$) since this is the most natural one tied to the Earth in describing 
the directional exposure of any experiment. The random sample $\{\mathbf{n}_1,\ldots,\mathbf{n}_N\}$ 
results from a Poisson process whose average is the flux of cosmic rays $\Phi(\mathbf{n})$ coupled to 
the directional exposure $\omega(\mathbf{n})$ of the considered experiment,
\begin{equation}
\label{eqn:avg}
\left\langle\frac{\mathrm{d}N(\mathbf{n})}{\mathrm{d}\Omega}\right\rangle=\omega(\mathbf{n})\,\Phi(\mathbf{n}).
\end{equation}
As for any angular distribution on the unit sphere, the flux of cosmic ray $\Phi(\mathbf{n})$ can be decomposed 
in terms of a multipolar expansion onto the spherical harmonics $Y_{\ell m}(\mathbf{n})$,
\begin{equation}
\label{eqn:almexpansion}
\Phi(\mathbf{n})=\sum_{\ell\geq0}\sum_{m=-\ell}^\ell a_{\ell m}Y_{\ell m}(\mathbf{n}).
\end{equation}
Any anisotropy fingerprint is encoded in the multipoles $a_{\ell m}$. Non-zero amplitudes in the $\ell$ modes 
contribute in variations of the flux on an angular scale ${\simeq}1/\ell$ radians. The rest of this section
is dedicated to the definition of an estimator $\bar{a}_{\ell m}$ of the multipolar coefficients and to the
derivation of the statistical properties of this estimator. 

With full-sky but non-uniform coverage, the customary recipe for decoupling directional exposure effects from 
anisotropy ones consists in weighting the observed angular distribution by the inverse of the \emph{relative} 
directional exposure function~\citep{sommers}
\begin{equation}
\label{eqn:obsfluxtilde}
\frac{\mathrm{d}\tilde{N}(\mathbf{n})}{\mathrm{d}\Omega}=\frac{1}{\bar{\omega}_r(\mathbf{n})}\frac{\mathrm{d}N(\mathbf{n})}{\mathrm{d}\Omega}.
\end{equation}
The relative directional exposure $\bar{\omega}_r$ is a dimensionless function normalized to unity at its maximum.
When the function $\omega$ (or $\omega_r$) is known from a single experiment, the averaged angular distribution 
$\langle\mathrm{d}\tilde{N}/\mathrm{d}\Omega\rangle$ is, from equation~\eqref{eqn:avg}, identified with the flux of cosmic
rays $\Phi(\mathrm{n})$ times the total exposure of the experiment. In turn, when combining the exposure of the two 
experiments, the relationship between $\langle\mathrm{d}\tilde{N}/\mathrm{d}\Omega\rangle$ and $\Phi(\mathrm{n})$ is no 
longer so straightforward due to the finite resolution in estimating the parameter $b$ introduced in 
equation~\eqref{eqn:omega}. To first order, it can be expressed as
\begin{equation}
\label{eqn:avg_obsfluxtilde}
\left\langle\frac{\mathrm{d}\tilde{N}(\mathbf{n})}{\mathrm{d}\Omega}\right\rangle\simeq\left\langle\frac{1}{\bar{\omega}_r(\mathbf{n})}\right\rangle\omega(\mathbf{n})\,\Phi(\mathbf{n}).
\end{equation}
For an unbiased estimator of $b$ with a resolution\footnote{The actual resolution 
in $b$ obtained in section~\ref{sec:jointmethod} is ${\simeq}3.9\%$.} not larger than ${\simeq}10\%$, the relative differences between $\langle1/\bar{\omega}_r(\mathbf{n})\rangle$ and $1/\omega_r(\mathbf{n})$ are actually not larger 
than $10^{-3}$ in such a way that $\langle\mathrm{d}\tilde{N}/\mathrm{d}\Omega\rangle$ can still be identified to 
$\Phi(\mathbf{n})$ times the total exposure to a very good approximation. Consequently, the $\bar{a}_{\ell m}$ 
coefficients recovered, defined as
\begin{equation}
\label{eqn:est_alm}
\bar{a}_{\ell m}=\int_{4\pi}\mathrm{d}\Omega\frac{\mathrm{d}\tilde{N}(\mathbf{n})}{\mathrm{d}\Omega}Y_{\ell m}(\mathbf{n})=\sum_{i=1}^N\frac{Y_{\ell m}(\mathbf{n}_i)}{\bar{\omega}_r(\mathbf{n}_i)},
\end{equation}
provide unbiased estimators of the underlying $a_{\ell m}$ multipoles since the relationship
$\langle\bar{a}_{\ell m}\rangle=a_{\ell m}$ can be established by propagating
equation~\eqref{eqn:avg_obsfluxtilde} into $\langle\bar{a}_{\ell m}\rangle$.

Using the estimators defined in equation~\eqref{eqn:est_alm}, the expected resolution $\sigma_{\ell m}$ 
on each $a_{\ell m}$ multipole can be inferred from the second moment of $\mathrm{d}\tilde{N}/\mathrm{d}\Omega$ 
accordingly to Poisson statistics,
\begin{equation}
\label{eqn:mom2_obsfluxtilde}
\left\langle\frac{\mathrm{d}\tilde{N}(\mathbf{n})}{\mathrm{d}\Omega}\frac{\mathrm{d}\tilde{N}(\mathbf{n}')}{\mathrm{d}\Omega'}\right\rangle=\left\langle\frac{1}{\bar{\omega}_r(\mathbf{n})\bar{\omega}_r(\mathbf{n}')}\right\rangle\Bigg[\omega(\mathbf{n})\omega(\mathbf{n}')\,\Phi(\mathbf{n})\,\Phi(\mathbf{n}')+\omega(\mathbf{n})\,\Phi(\mathbf{n})\,\delta(\mathbf{n},\mathbf{n}')\Bigg].
\end{equation}

\begin{figure}[!t]
  \centering					 
 \includegraphics[width=0.7\textwidth]{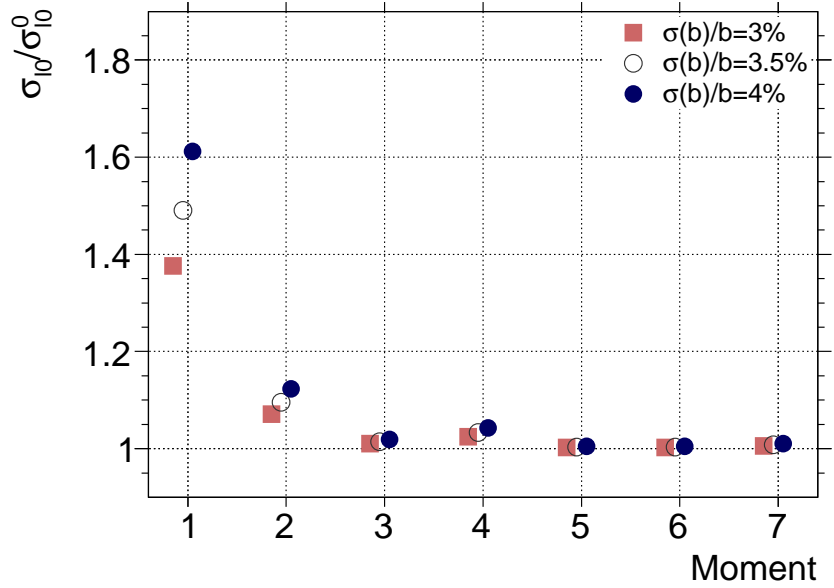}
  \caption{\small{Influence of the uncertainty in the relative exposures between the two experiments on 
  the resolution of the recovered $\bar{a}_{\ell 0}$ coefficients, for different values of the resolution 
  on $b$. On the $y$-axis, the term $\sigma_{\ell 0}^0$ is obtained by dropping the second term inside 
  the square root in the expression of $\sigma_{\ell 0}$ (see equation~\eqref{eqn:rms_alm}).}}
  \label{fig:degradation}
\end{figure}

Once propagated into the covariance matrix of the estimated $\bar{a}_{\ell m}$ coefficients, 
equation~\eqref{eqn:mom2_obsfluxtilde} allows the determination of $\sigma_{\ell m}$ in the case
of relatively small $\{a_{\ell m}\}_{\ell\geq 1}$ coefficients compared to $a_{00}$. Using as above
the fact that $\langle1/\bar{\omega}_r(\mathbf{n})\rangle$ can be accurately replaced
by $1/\omega_r(\mathbf{n})$, the resolution parameters $\sigma_{\ell m}$ read
\begin{align}
\label{eqn:rms_alm}
\sigma_{\ell m}\simeq\Bigg[&\frac{a_{00}}{\sqrt{4\pi}}\int_{4\pi}\mathrm{d}\Omega\left\langle\frac{1}{\bar{\omega}_r^2(\mathbf{n})}\right\rangle\omega(\mathbf{n})\,Y_{\ell m}^2(\mathbf{n})+ \nonumber\\
&\frac{a_{00}^2}{4\pi}\int_{4\pi}\mathrm{d}\Omega\,\mathrm{d}\Omega'
\left[\left\langle\frac{1}{\bar{\omega}_r(\mathbf{n})\,\bar{\omega}_r(\mathbf{n}')}\right\rangle\omega(\mathbf{n})\,\omega(\mathbf{n}')-1\right]Y_{\ell m}(\mathbf{n})\,Y_{\ell m}(\mathbf{n}')\Bigg]^{1/2}.
\end{align}
If $b$ was known with perfect accuracy, the second term in equation~\eqref{eqn:rms_alm} would vanish, and 
the resolution of the $a_{\ell m}$ coefficients would be driven by Poisson fluctuations only as in the case of a single 
experiment. But, having at one's disposal an estimation of $b$ only, the second term reflects the effect of the 
uncertainty in the relative exposures of the two experiments. For a directional exposure independent of the right 
ascension, the azimuthal dependences of the spherical harmonics can be factorized from the whole solid angle
integrations so that the whole term is non-zero only for $m=0$. Its influence is illustrated in figure~\ref{fig:degradation}, 
where the ratio between the total expression of $\sigma_{\ell 0}$ and the partial one, ignoring this second term 
inside the square root, is plotted as a function of the multipole $\ell$ for different resolution values on $b$. While 
this ratio amounts to ${\simeq}1.5$ for $\ell=1$ and $\sigma(b)/b=3.5\%$, it falls to ${\simeq}1.1$ for $\ell=2$ and 
then tends to 1 for higher multipole values. Consequently, in accordance with naive expectations, the uncertainty 
in the $b$ factor mainly impacts the resolution in the dipole coefficient $a_{10}$ while it has a small influence 
on the quadrupole coefficient $a_{20}$ and a marginal one on higher order moments $\{a_{\ell 0}\}_{\ell\geq3}$.

\section{The Joint-Analysis Method and its Performances}
\label{sec:jointmethod}

A band of declinations around the equatorial plane is exposed to the fields of view of both experiments,
namely for declinations between $-15^\circ$ and $25^\circ$. This overlapping region can be used for 
designing an empirical procedure to get a relevant estimate of the parameter $b$. The basic starting point
is the following. For an isotropic flux, the fluxes measured independently by both experiments in the common 
band would have to be identical. The commonly covered declination band could thus be used for \emph{cross-calibrating 
empirically the fluxes} of the experiments and for delivering an overall unbiased estimate of the parameter $b$. 
Since the shapes of the exposure functions are not identical in the overlapping region (see figure~\ref{fig:exposure}), 
the observed fluxes are not expected to be identical in case of anisotropies. For small anisotropies however, this 
guiding idea can nevertheless be implemented in an iterative algorithm delivering finally estimates of the 
parameter $b$ and of the multipole coefficients at the same time. 

Let us consider a joint data set with all events detected in excess of some energy threshold. The way the 
individual energy scales are chosen to select a common threshold, be it by using nominal energies or any 
cross-calibration procedure, does not matter at this stage. For anisotropies which do not vary suddenly with
energy, only a reasonable starting point is required to
guarantee that the anisotropy search pertains to events with energies in excess of roughly the same energy
threshold for both experiments. Then, considering as a first approximation the flux $\Phi(\mathbf{n})$ as 
isotropic, the overlapping region $\Delta\Omega$ can be utilized to derive a first estimate $\bar{b}^{(0)}$ 
of the $b$ factor by \emph{forcing the fluxes of both experiments to be identical in this particular region}.
This can be easily achieved in practice, by taking the ratio of the $\Delta N_\text{TA}$ and 
$\Delta N_\text{Auger}$ events observed in the overlapping region $\Delta\Omega$ weighted by 
the ratio of nominal exposures,
\begin{equation}
\label{eqn:b0}
\bar{b}^{(0)}=\frac{\Delta N_\text{Auger}}{\Delta N_\text{TA}}\,\frac{\displaystyle\int_{\Delta\Omega}\mathrm{d}\Omega\,\omega_\text{TA}(\mathbf{n})}{\displaystyle\int_{\Delta\Omega}\mathrm{d}\Omega\,\omega_\text{Auger}(\mathbf{n})}.
\end{equation}
Then, inserting $\bar{b}^{(0)}$ into $\bar{\omega}$, 'zero-order' $\bar{a}_{\ell m}^{(0)}$ 
coefficients can be obtained. This set of coefficients is only a rough estimation, due to the limiting assumption 
on the flux (isotropy).

\begin{figure}[!t]
  \centering					 
 \includegraphics[width=0.48\textwidth]{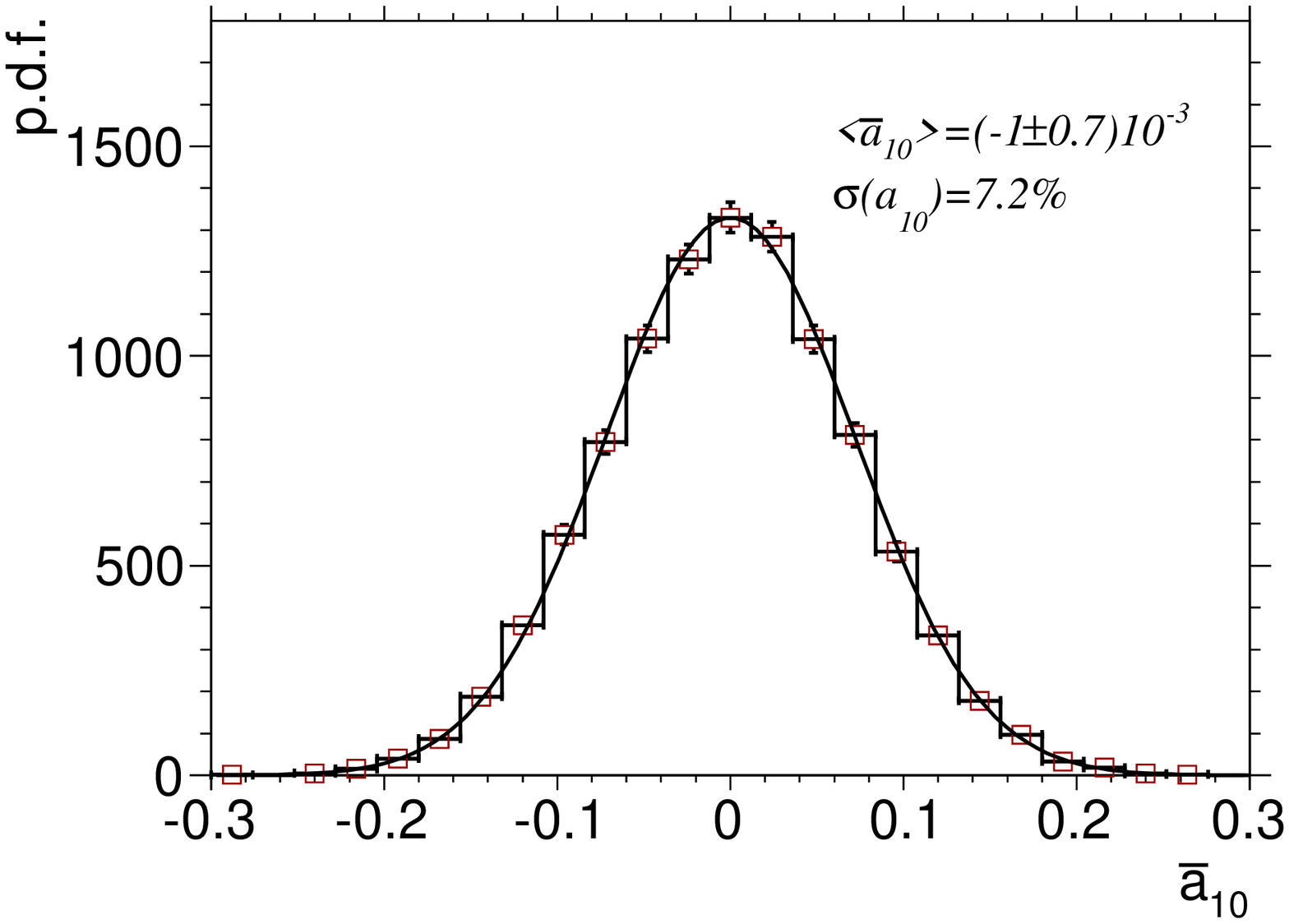}
 \includegraphics[width=0.48\textwidth]{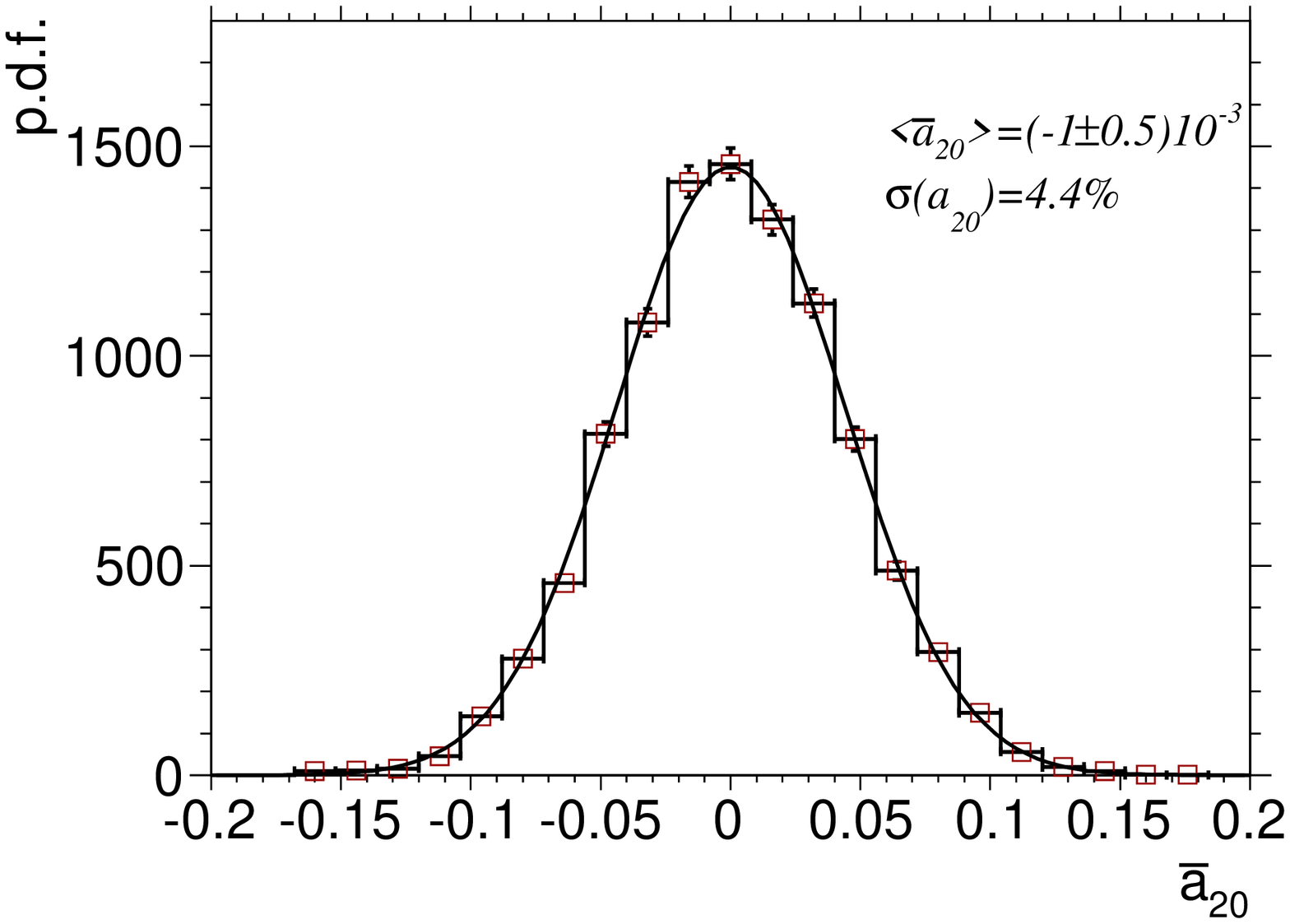}
  \caption{Reconstruction of $a_{10}$ (left) and $a_{20}$ (right) with the iterative procedure, in the case 
  of an underlying isotropic flux. Expectations are shown as the Gaussian curves whose resolution parameters 
  are from equation~\eqref{eqn:rms_alm}.}
  \label{fig:mciso}
\end{figure}

\begin{figure}[!t]
  \centering					 
 \includegraphics[width=0.48\textwidth]{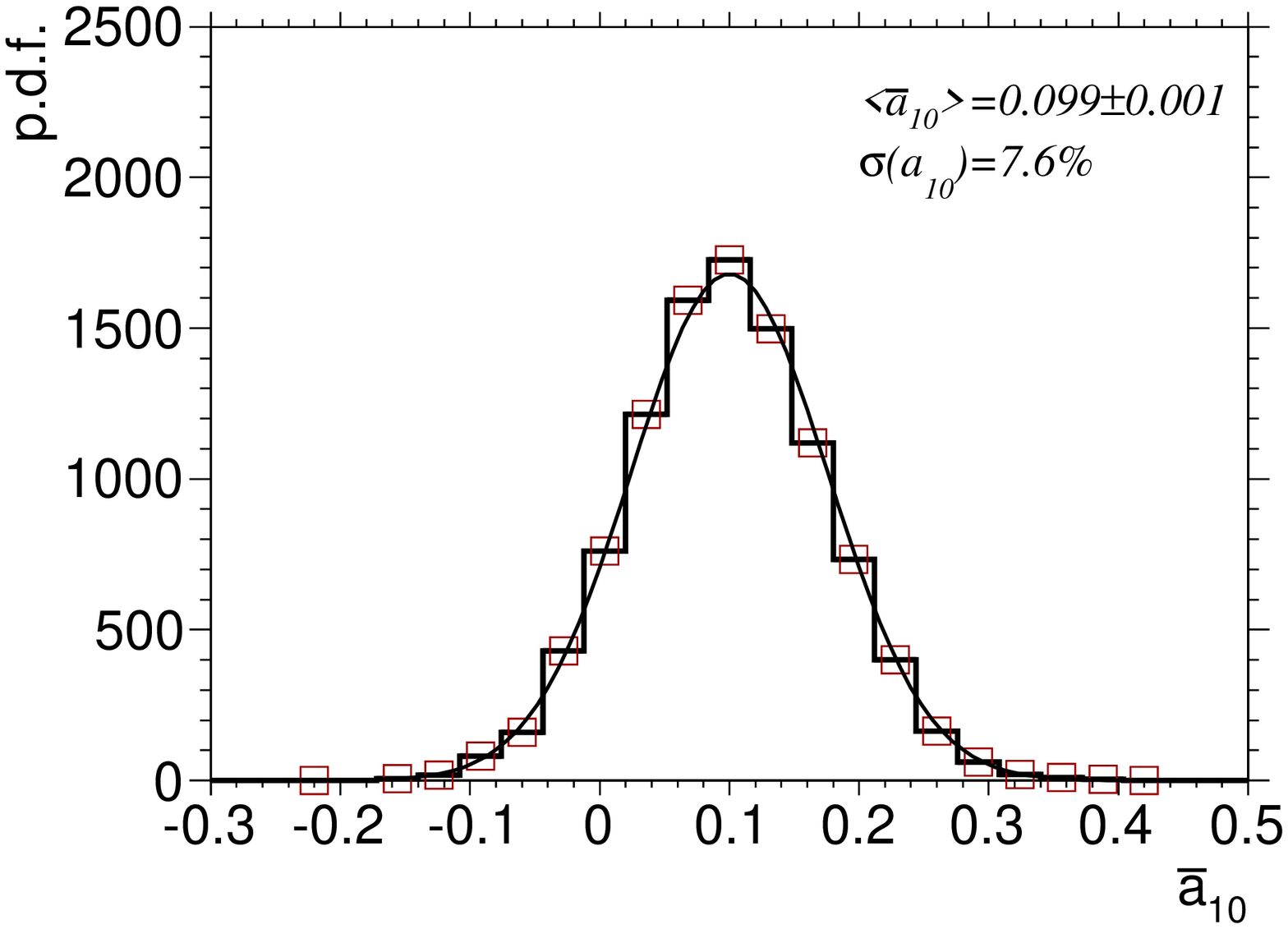}
 \includegraphics[width=0.48\textwidth]{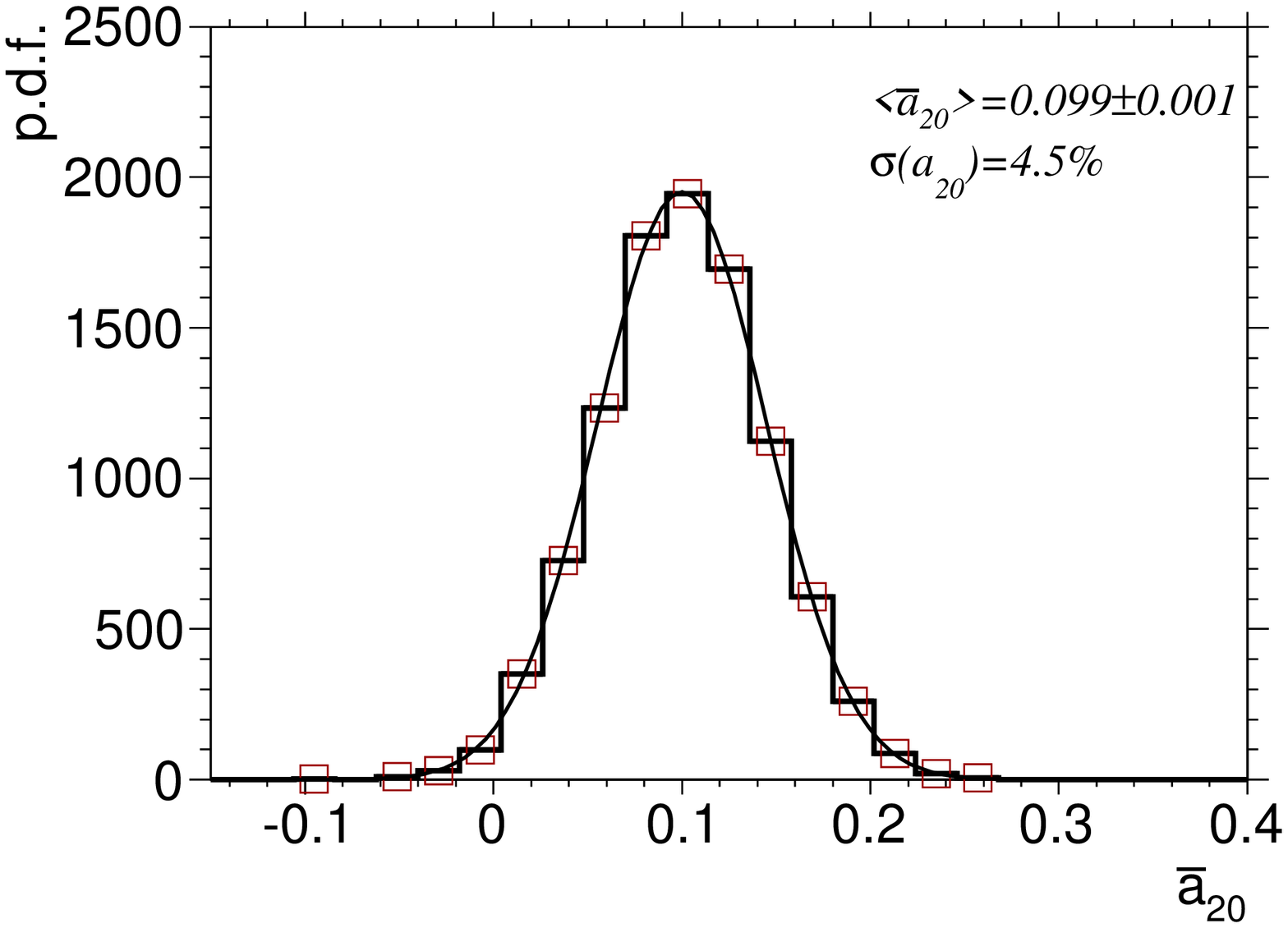}
  \caption{Same as figure~\ref{fig:mciso}, in the case of an anisotropic input flux 
  $\Phi(\mathbf{n})\propto1+0.1\,Y_{10}(\mathbf{n})+0.1\,Y_{20}(\mathbf{n})$.}
  \label{fig:mcanis}
\end{figure}\

On the other hand, the expected number of events in the common band for each observatory,
$\Delta n_\text{TA}^\text{exp}$ and $\Delta n_\text{Auger}^\text{exp}$, can be 
expressed from the underlying flux $\Phi(\mathbf{n})$ and the true value of $b$ as
\begin{align}
\label{eqn:dn}
\Delta n_\text{TA}^\text{exp}&=\int_{\Delta\Omega}\mathrm{d}\Omega\,\Phi(\mathbf{n})\,\omega_\text{TA}(\mathbf{n}) \nonumber \\
\Delta n_\text{Auger}^\text{exp}&=b\int_{\Delta\Omega}\mathrm{d}\Omega\,\Phi(\mathbf{n})\,\omega_\text{Auger}(\mathbf{n}).
\end{align}
From equations~\eqref{eqn:dn}, and from the set of $\bar{a}_{\ell m}^{(0)}$ coefficients, an iterative procedure 
estimating at the same time $b$ and the set of $a_{\ell m}$ coefficients can be constructed as
\begin{equation}
\label{eqn:b_est}
\bar{b}^{(k+1)}=\frac{\Delta N_\text{Auger}}{\Delta N_\text{TA}}\,\frac{\displaystyle\int_{\Delta\Omega}\mathrm{d}\Omega\,\bar{\Phi}^{(k)}(\mathbf{n})\,\omega_\text{TA}(\mathbf{n})}{\displaystyle\int_{\Delta\Omega}\mathrm{d}\Omega\,\bar{\Phi}^{(k)}(\mathbf{n})\,\omega_\text{Auger}(\mathbf{n})},
\end{equation}
where $\Delta N_\text{TA}$ and $\Delta N_\text{Auger}$ as derived in the first step are used 
to estimate $\Delta n_\text{TA}^\text{exp}$ and $\Delta n_\text{Auger}^\text{exp}$ 
respectively, and $\bar{\Phi}^{(k)}$ is the flux estimated with the set of $\bar{a}_{\ell m}^{(k)}$ 
coefficients. 

Whether this iterative procedure finally delivers unbiased estimations of the set of $a_{\ell m}$ 
coefficients with a resolution obeying equation~\eqref{eqn:rms_alm} can be tested by Monte-Carlo simulations. 
10\,000 mock samples are used here, with a number of events similar to the one of the actual joint data set and 
with ingredients corresponding to the actual figures in terms of total and directional exposures. Under these
realistic conditions, the resolution obtained on the $b$ parameter is found to be ${\simeq}3.9\%$. The distributions 
of the reconstructed low-order $\bar{a}_{10}$ and $\bar{a}_{20}$ multipole coefficients, which are 
\emph{a priori} the most challenging to recover, are shown in figure~\ref{fig:mciso} after $k=10$ iterations in the 
case of an underlying isotropic flux of cosmic rays. The reconstructed histograms are observed to be well-described 
by Gaussian functions centered on zero and with a dispersion following indeed equation~\eqref{eqn:rms_alm}. 

With exactly the same ingredients, the simulations can be used to test the procedure with an underlying anisotropic 
flux of cosmic rays, chosen here such that $\Phi(\mathbf{n})\propto1+0.1\,Y_{10}(\mathbf{n})+0.1\,Y_{20}(\mathbf{n})$. 
Results of the Monte-Carlo simulations are shown in figure~\ref{fig:mcanis} for the specific $a_{10}$ 
and $a_{20}$ coefficients. Again, the reconstructed histograms are observed to be well described by Gaussian functions
with parameters following the expectations.

Note that in practice, all results presented here are found to be stable as soon as the number of iterations is $k=4$. 

Formally, the implementation of the cross-calibration procedure is not limited to the choice of the whole overlapping
declination band for the integration range $\Delta\Omega$ in previous equations. The choice of the whole common 
band turns out to be, however, optimal in terms of resolution in $b$. A restriction of the common declination band to, 
for instance, $[-10^\circ,10^\circ]$ would lead to a resolution in $b$ of ${\simeq}5\%$; while the use of the whole sky 
would not bring any improvement for resolving better $b$. In next sections, the cross-calibration procedure is thus 
applied to the joint data set by using the whole overlapping region for $\Delta\Omega$.

\section{Joint Data Analysis}
\label{sec:dataanalysis}

\begin{figure}[!t]
  \centering					 
 \includegraphics[width=0.9\textwidth]{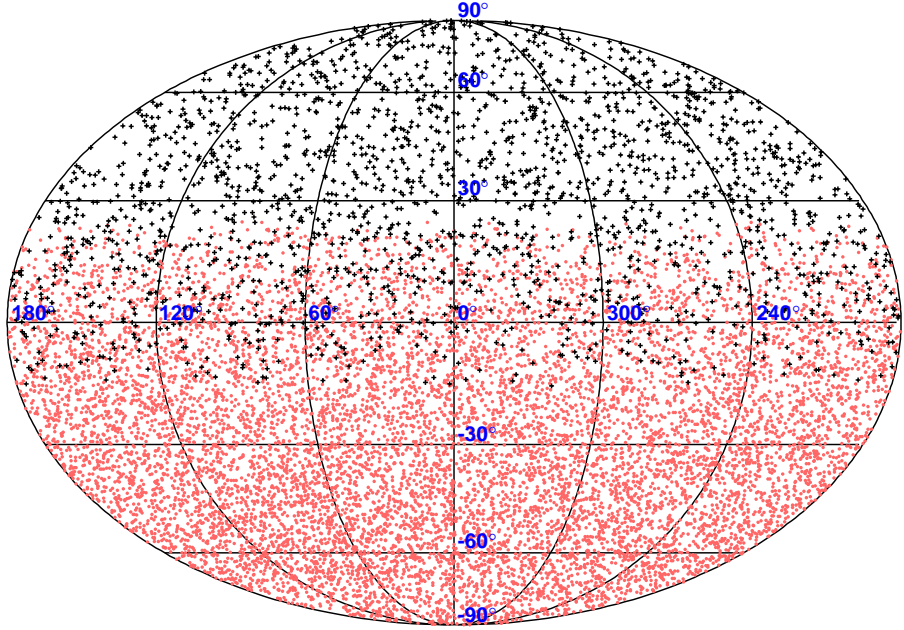}
  \caption{Arrival directions of Auger events (red points in the South hemisphere) and Telescope Array ones
  (black crosses in the Northern hemisphere) above $10^{19}$\,eV in equatorial coordinates, using a Mollweide 
  projection.}
  \label{fig:rawdata}
\end{figure}

All analyses reported in this section are based on a joint data set consisting of events with energies in excess of 
roughly $10^{19}$\,eV in terms of the energy scale used at the Telescope Array by evaluating in the Auger data set 
the energy threshold which guarantees equal fluxes for both experiments. We are thus left here with 2130 events
(795 in the common band) above $10^{19}$\,eV from the Telescope Array and 11\,087 (3435 in the common band) above 
$8.5{\times}10^{18}$\,eV from the Pierre Auger Observatory. The arrival directions are shown in figure~\ref{fig:rawdata}
in equatorial coordinates using the Mollweide projection. Auger data can be seen as the red points in the Southern
hemisphere, while Telescope Array ones are shown as the black crosses in the Northern hemisphere.

The methodology presented in the previous section allows us to estimate the multipole coefficients and to
perform a rich series of anisotropy searches by taking profit of the great advantage offered by the full-sky coverage. 
After iterations, the coefficient $b$ is $b=1.011$.
Choosing to use nominal energies to build the joint data set would lead to a different value for $b$ (0.755) due to the
different statistics in the Auger data set (8259 events in total instead of 11\,087),
but it will be shown in next section that this choice impacts the physics results to only a small extent.

\begin{figure}[!t]
  \centering					 
 \includegraphics[width=0.48\textwidth]{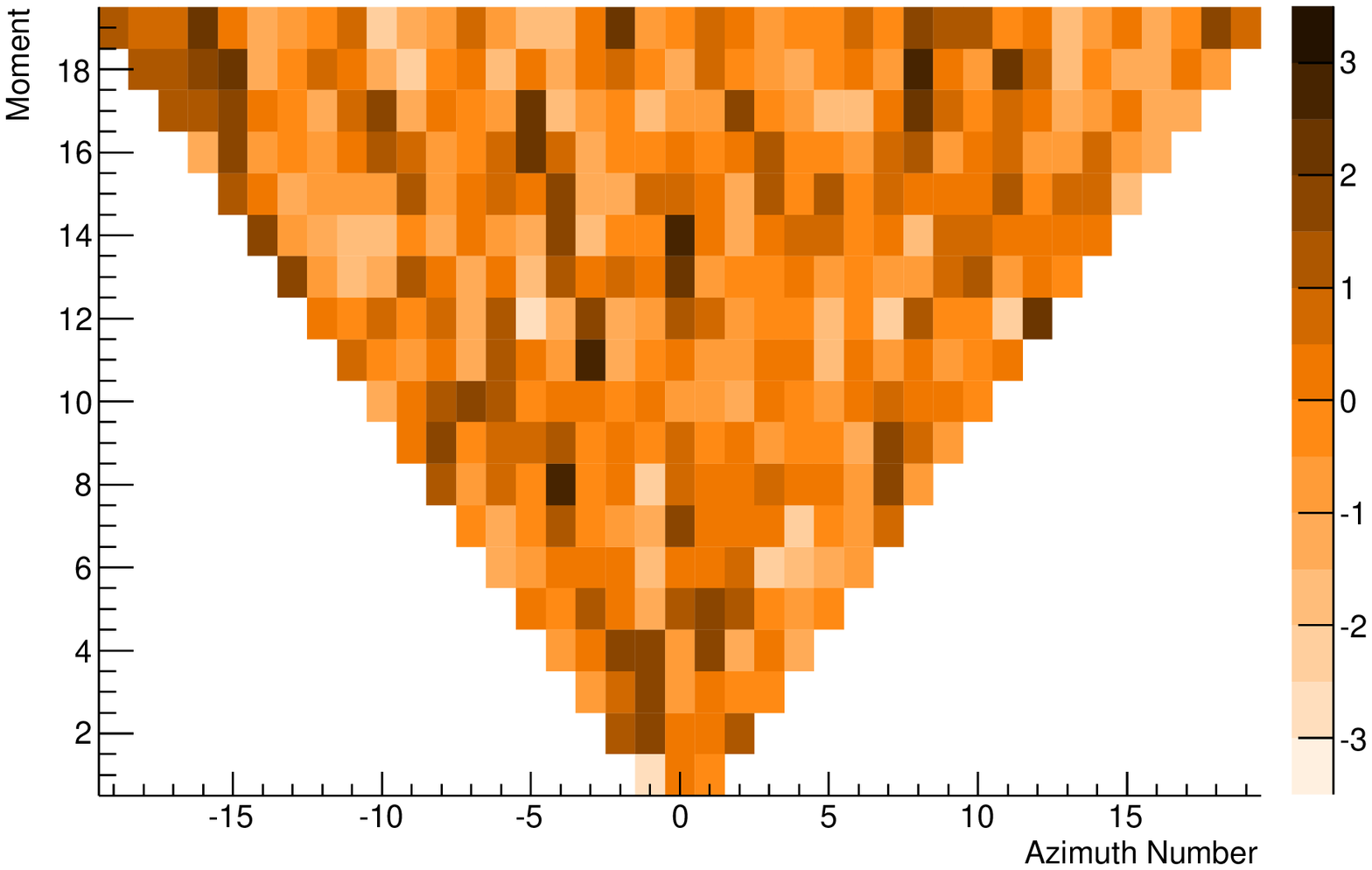}
 \includegraphics[width=0.43\textwidth]{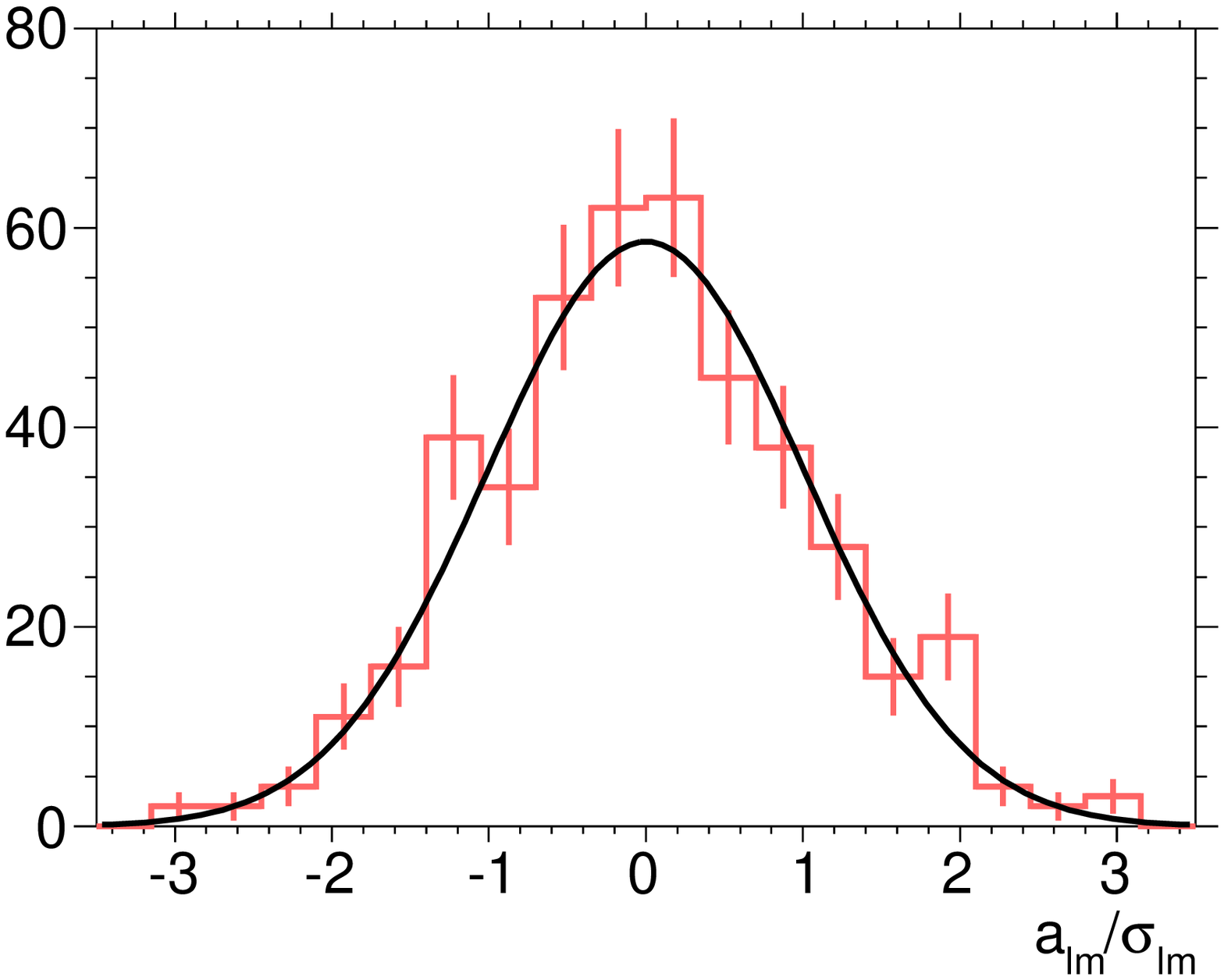}
  \caption{Significance table (left) and histogram (right) of the estimated multipole moments (in equatorial coordinates).
  In the right panel, the black line is a normal curve.}
  \label{fig:alm_sigtable}
\end{figure}

The normalization convention of the multipole moments used hereafter is chosen so that the $a_{\ell m}$
coefficients measure the relative deviation with respect to the whole contribution of the monopole (i.e.\ the 
$a_{\ell m}$ coefficients are redefined such that $a_{\ell m} \rightarrow \sqrt{4\pi}a_{\ell m}/a_{00}$).

\subsection{Multipolar Analysis}
\label{sec:multipoles}

\begin{table}[!h]
\begin{center}
\begin{tabular}{rrrrrrrrrrr}
\hline\hline
\multicolumn{1}{l}{$\ell$} & \multicolumn{1}{l}{$m$} & \multicolumn{1}{l}{$a_{\ell m}$} & ~~ & \multicolumn{1}{l}{$\ell$} & \multicolumn{1}{l}{$m$} & \multicolumn{1}{l}{$a_{\ell m}$} & ~~ & \multicolumn{1}{l}{$\ell$} & \multicolumn{1}{l}{$m$} & \multicolumn{1}{l}{$a_{\ell m}$}  \\
\cline{1-3} \cline{5-7} \cline{9-11}
   &     &        &     & & & & &   &  $-3$ & $-0.022\pm0.034$ \\
   &     &    &    & & $-2$ & $0.038\pm0.035$ &  &  &  $-2$ & $0.030\pm0.039$ \\
   &  $-1$ & $-0.102\pm0.036$ & & & $-1$ & $0.067\pm0.040$ & & & $-1$ & $0.067\pm0.037$ \\
 1 &  0 & $0.006\pm0.074$ & & 2 & 0 & $0.017\pm0.042$ & & 3 &  0 & $-0.027\pm0.040$ \\
   &  1 & $-0.001\pm0.036$ & & &  1 & $0.004\pm0.040$ & &   &  1 & $0.009\pm0.037$ \\
   &     &    &  &  &  2 & $0.040\pm0.035$ &  &   &  2 & $-0.004\pm0.039$ \\
   &     &  &      &     & & & &    &  3 & $-0.011\pm0.034$ \\
\hline
\end{tabular}
\caption{First low-order multipolar moments and their uncertainties (in equatorial coordinates).}
\label{tab:alm}
\end{center}
\end{table}

The dipole, quadrupole and octupole moments as derived from the iterative procedure are reported in
table~\ref{tab:alm} in equatorial coordinates together with their associated uncertainties calculated from
equation~\eqref{eqn:rms_alm}. None of these multipole coefficients stands out as being significantly above 
the noise level. 

The full set of multipole coefficients provides a comprehensive description of the anisotropy patterns that might
be present in the data. A significance table for the coefficients up to $\ell=20$, built simply by dividing each
estimated coefficient by its corresponding uncertainty, is reported in the left panel of figure~\ref{fig:alm_sigtable}. 
As it can be seen from the contrast scale, significance values between $-1$ and 1 dominate the picture. Deviations
close to $-3$ and 3 stand at the expected level for isotropy, as shown in the right panel. Hence, overall, the extraction 
of the multipole coefficients does not provide any evidence for anisotropy.

\subsection{Flux and Overdensities/Underdensities Sky Maps}
\label{sec:maps}

\begin{figure}[!t]
  \centering					 
 \includegraphics[width=0.48\textwidth]{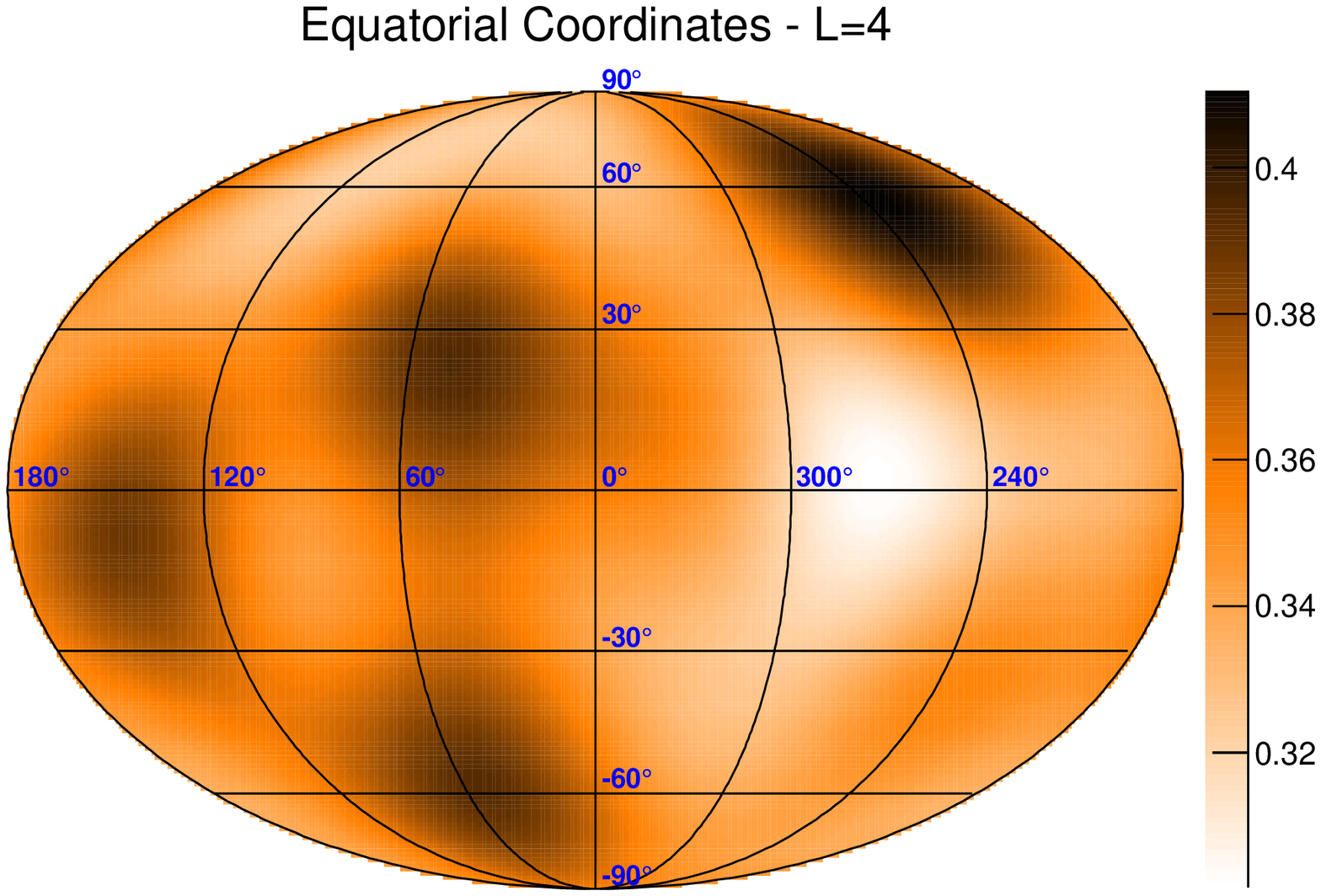}
 \includegraphics[width=0.48\textwidth]{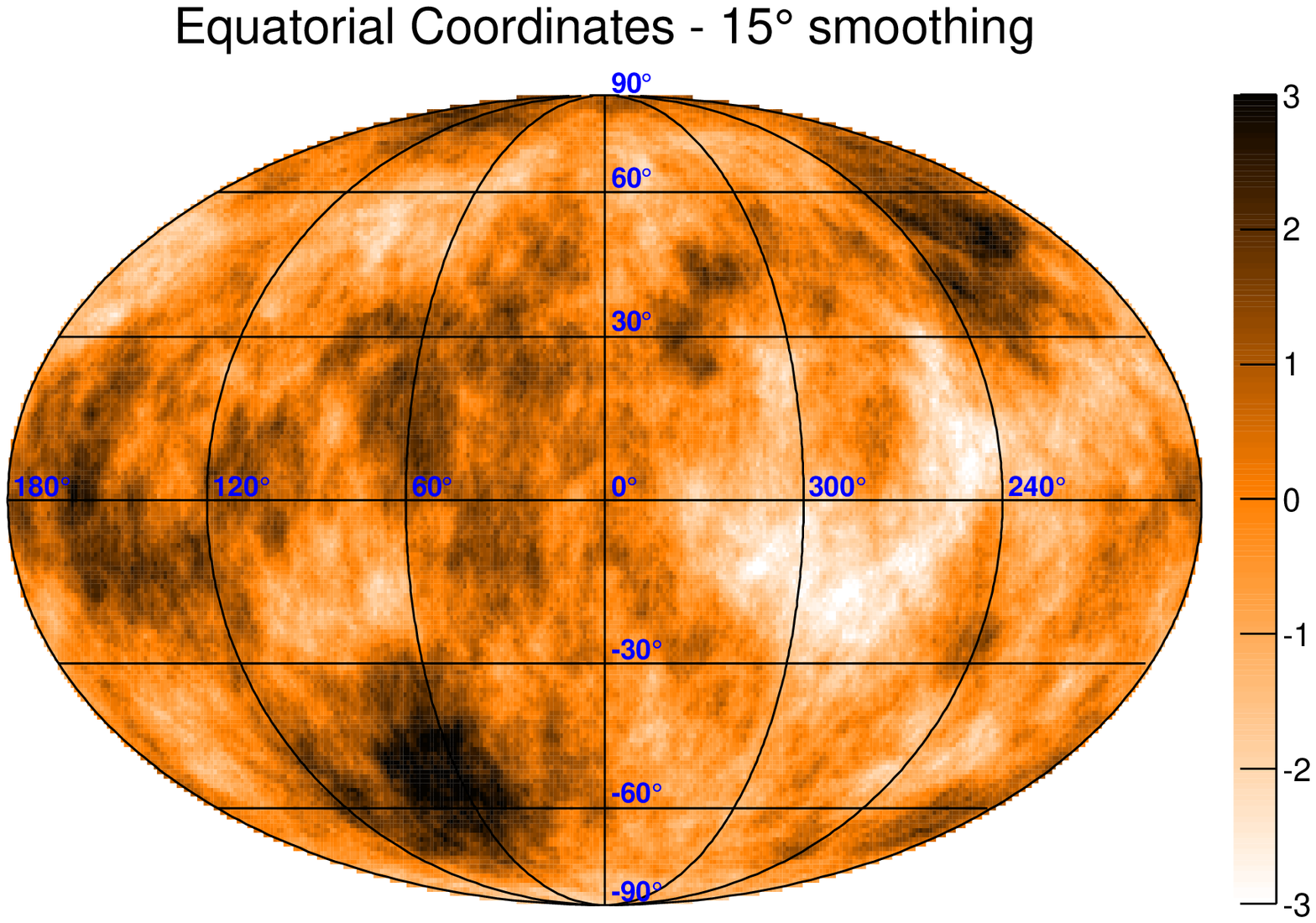}
  \caption{Left: Flux sky map in km$^{-2}$yr$^{-1}$sr$^{-1}$ units, using a multipolar expansion up to $\ell=4$. Right: Significance sky map smoothed out at a $15^\circ$ angular scale.}
  \label{fig:maps}
\end{figure}

To visualise the result of the multipolar expansion, a flux sky map of the joint data set is displayed using the Mollweide
projection in the left panel of figure~\ref{fig:maps}, in km$^{-2}$yr$^{-1}$sr$^{-1}$ units. This map is drawn in equatorial
coordinates. To exhibit structures at intermediate scales, the expansion is truncated here at $\ell=4$. Relative excesses
and deficits are clearly visible on a 15\% contrast scale.

To quantify whether some contrasts are statistically compelling or not, a significance sky map of the overdensities/underdensities
obtained in circular windows of radius 15$^\circ$ is shown in the right panel. The choice of the 15$^\circ$ angular scale is
well suited to exhibit structures at scales that can be captured by the set of low-order multipoles up to $\ell=4$. Significances
are calculated using the widely used Li and Ma estimator~\citep{lima}, $S$, which was designed to account for both the 
fluctuations of the background and of an eventual signal in any angular region searched,
\begin{equation}
\label{eqn:lima}
S=\pm\sqrt{2}\left[N_\text{on}\ln{\frac{(1+\alpha_\text{LM})N_\text{on}}{\alpha_\text{LM}(N_\text{on}+N_\text{off})}}+N_\text{off}\ln{\frac{(1+\alpha_\text{LM})N_\text{off}}{N_\text{on}+N_\text{off}}}\right]^{1/2},
\end{equation}
with $N_\text{on}$ the observed number of events in the angular region searched, and $N_\text{off}$ the
\emph{total} number of events. The sign of $S$ is chosen positive in case of excesses and negative in case of deficits. On
the other hand, since the background estimation is not based here on any \emph{on/off} procedure but can be instead
determined from the integration of the directional exposure in the angular region searched, the $\alpha_\text{LM}$
parameter expressing the expected ratio of the count numbers between the angular region searched and any background
region is taken here as
\begin{equation}
\label{eqn:alpha_lima}
\alpha_\text{LM}(\mathbf{n})=\frac{\displaystyle\int\mathrm{d}\mathbf{n}'\,\bar{\omega}(\mathbf{n}')\,f(\mathbf{n},\mathbf{n}')}{\displaystyle\int\mathrm{d}\mathbf{n}\,\bar{\omega}(\mathbf{n})},
\end{equation}
with $f$ the top-hat filter function at the angular scale of interest. In absence of signal, the variable $S$ is expected
to be nearly normally distributed. Hence, for positive (negative) values, $S$ ($-S$) can be interpreted as the number of
standard deviations of any excess (deficit) in the sky.

Overall, overdensities and underdensities obtained in circular windows of radius 15$^\circ$ are well reproduced by the 
multipolar expansion. Contrasts are not identical in all regions of the sky due to the non-uniform coverage (high flux values 
in low-exposed regions can lead to overdensities less significant than lower flux values in higher exposed regions, and 
\emph{vice-versa}), but the overall pattern looks similar between the two maps. From the significance contrast scale in the 
right panel, it is clear that there is no overdensity or underdensity standing above the 3 standard deviation level. The 
distribution of significances turns out to be compatible with that expected from fluctuations of an isotropic distribution.

\subsection{Dipole and Quadrupole Moments}
\label{sec:dipquad}

As outlined in the introduction, although the full set of spherical harmonic moments is needed to characterise any
departure from isotropy at any scale, the dipole and quadrupole moments are of special interest. For that reason, 
a special emphasis is given here to these low-order moments, in terms of a more traditional and geometric representation 
than the raw result of the multipole moments. 

\begin{figure}[!t]
  \centering					 
 \includegraphics[width=0.48\textwidth]{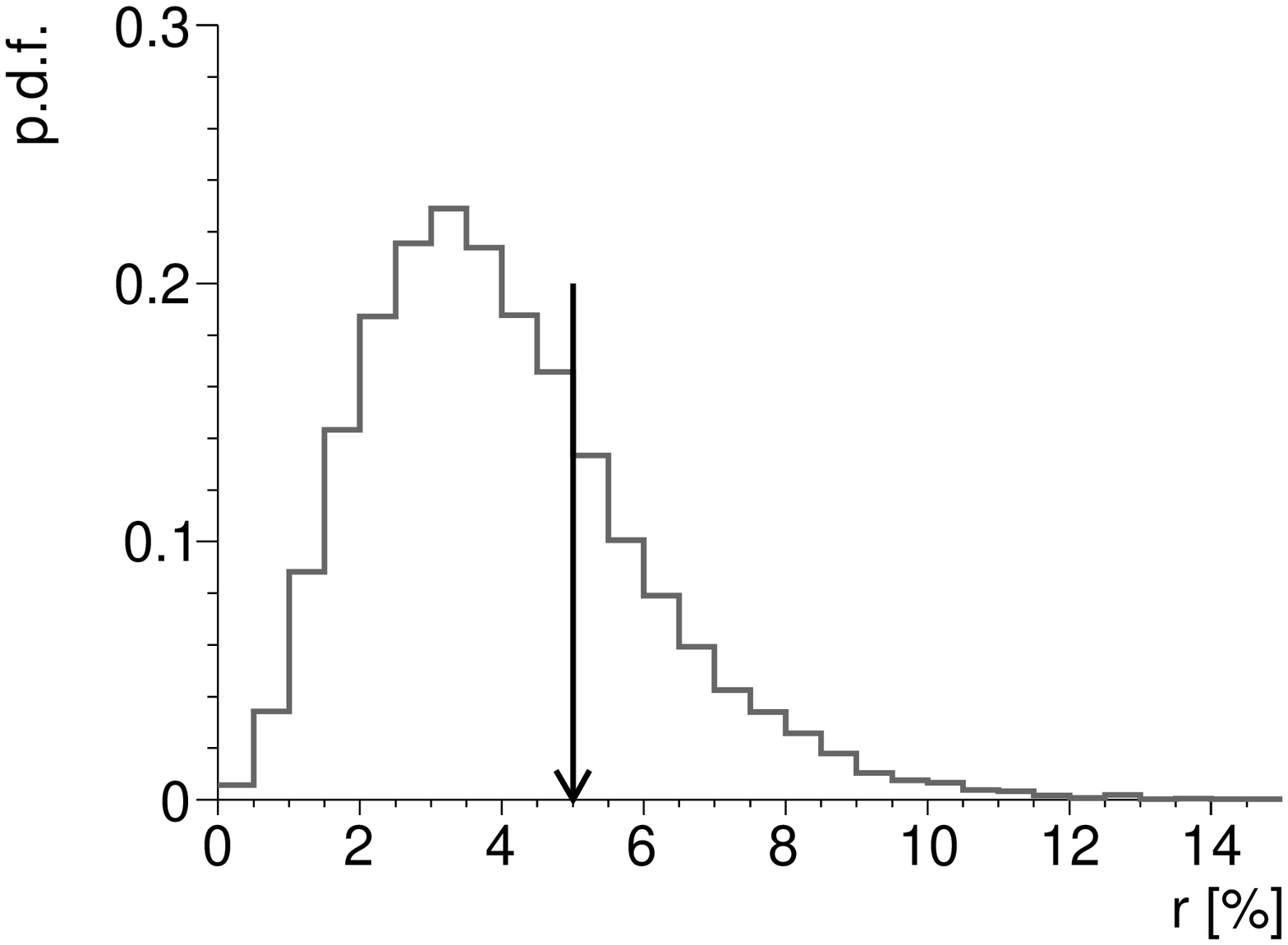}
 \includegraphics[width=0.48\textwidth]{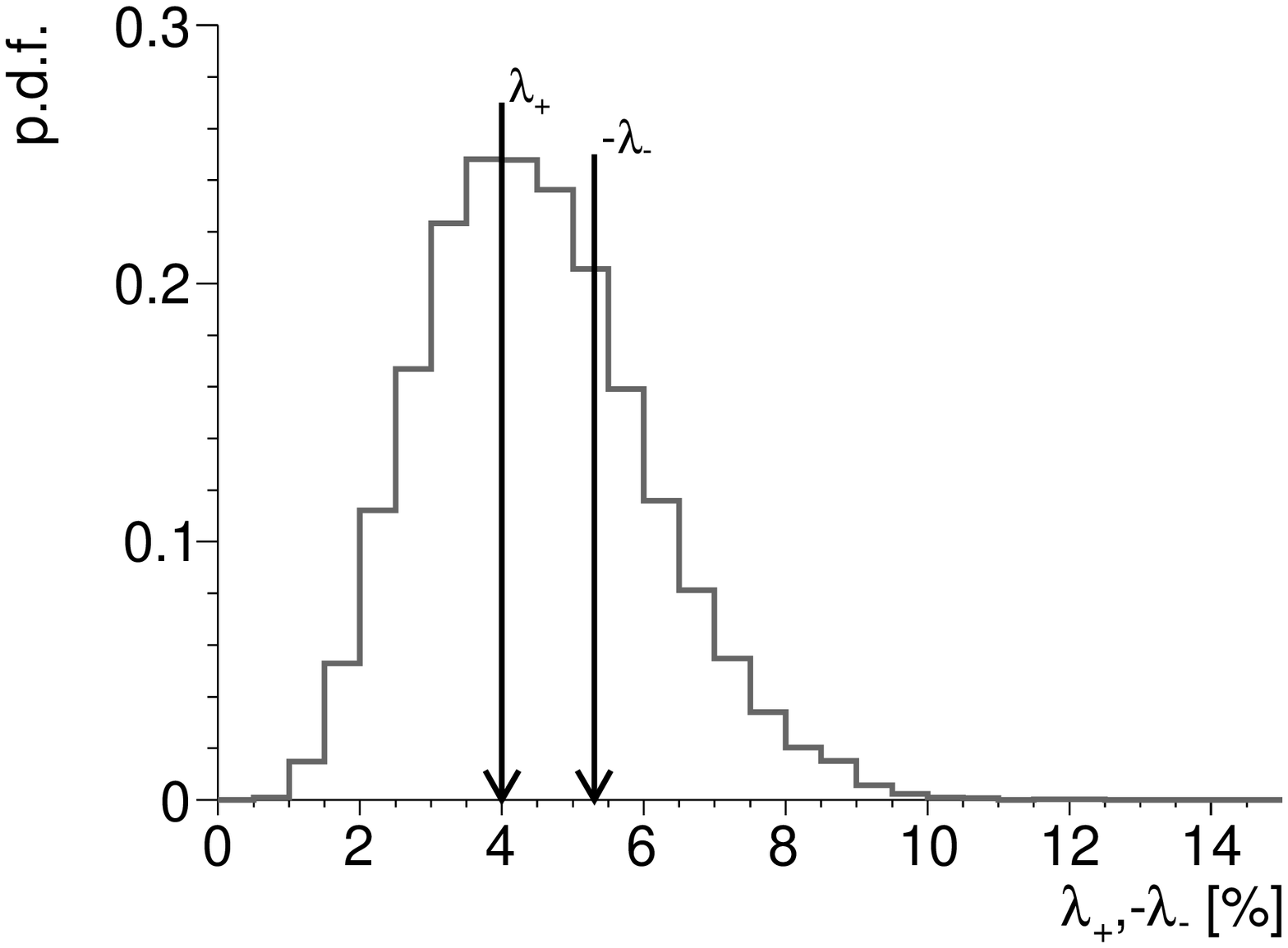}
  \caption{Measured amplitudes for the dipole vector (left) and the quadrupole tensor (right), together with
  the distributions expected from statistical fluctuations of isotropy.}
  \label{fig:dip_quad}
\end{figure}

The dipole moment can be fully characterized by a vector with an amplitude $r$ and the two angles $\{\delta_\text{d},\alpha_\text{d}\}$
of the unit vector $\mathbf{d}$. The quadrupole, on the other hand, can be fully determined by two independent amplitudes
$\{\lambda_+,\lambda_-\}$, two angles $\{\delta_{\text{q}_+},\alpha_{\text{q}_+}\}$ defining the orientation of a unit vector $\mathbf{q}_+$,
and one additional angle $\alpha_{q_-}$ defining the directions of another unit vector $\mathbf{q_-}$ in the orthogonal plane to
$\mathbf{q}_+$. The full description is completed by means of a third unit vector $\mathbf{q}_0$, orthogonal to both
$\mathbf{q}_+$ and $\mathbf{q}_-$, and with a corresponding amplitude such that the traceless condition
$\lambda_++\lambda_0+\lambda_-=0$ is satisfied. The estimation of the amplitudes and angles of the unit vectors
from the estimated spherical harmonic moments is straightforward (see appendix). The parameterisation of the low-order moments of the flux
is then written in a convenient and intuitive way as
\begin{equation}
\label{eqn:phidipquad}
\Phi(\mathbf{n})=\frac{\Phi_0}{4\pi}\left(1+r\,\mathbf{d}\cdot\mathbf{n}+\lambda_+(\mathbf{q}_+\cdot\mathbf{n})^2
+\lambda_0(\mathbf{q}_0\cdot\mathbf{n})^2+\lambda_-(\mathbf{q}_-\cdot\mathbf{n})^2+\cdots\right).
\end{equation}

\begin{table}[!t]
\begin{center}
\begin{tabular}{lrrrrr}
\hline\hline
 & \multicolumn{1}{l}{amplitude [\%]} & \multicolumn{1}{l}{$\delta [^\circ]$} & \multicolumn{1}{l}{$\alpha [^\circ]$} & \multicolumn{1}{l}{$l [^\circ]$} & \multicolumn{1}{l}{$b [^\circ]$} \\
\hline
$\mathbf{d}$   &  $5.0\pm1.8$ &   $3\pm30$ &  $89\pm22$ & 204 & $-10$ \\
$\mathbf{q}_+$ &  $4.0\pm1.8$ & $-42\pm41$ &  $46\pm69$ & 260 & 58 \\
$\mathbf{q}_-$ & $-5.3\pm2.0$ &  $28\pm22$ & $106\pm76$ & 154 & 25 \\
$\mathbf{q}_0$ &  $1.3\pm1.6$ &  $34\pm29$ & $354\pm72$ & 113 & $-5$ \\
\hline
\end{tabular}
\caption{Amplitudes and angles of the dipole vector and quadrupole tensor.}
\label{tab:dipquad}
\end{center}
\end{table}

The values of the estimated amplitudes and angles are given in table~\ref{tab:dipquad} with their associated
uncertainties. The distributions of amplitudes obtained from statistical fluctuations of simulated isotropic samples
are shown in figure~\ref{fig:dip_quad}. The superimposed arrows, indicating the measured values, are clearly
seen to stand within high probable ranges of amplitudes expected from isotropy. 

The dipole parameters, namely the amplitude, declination and phase, are observed to be compatible with 
previous reports from both experiments~\citep{auger-ls,auger-ls3d,auger-lsl,uhecr12}. It is worth noting this for
the phase $\alpha_\text{d}$ of the dipole vector $\mathbf{d}$: a consistency of phases in adjacent
energy intervals was also pointed out in the Auger data~\citep{auger-ls,auger-ls3d}. Given that a consistency of phases
is expected to manifest with a smaller number of events than those required for the detection of significant
amplitudes, continued scrutiny of future data will provide evidences of whether this phase consistency in 
both hemispheres is indicative of a real anisotropy or not.

\subsection{Power Spectrum}
\label{sec:powerspectrum}

The angular power spectrum $C_\ell$ is a coordinate-independent quantity, defined as the average 
$\left|a_{\ell m}\right|^2$ as a function of $\ell$,
\begin{equation}
\label{eqn:cell}
C_\ell=\frac{1}{2\ell+1}\sum_{m=-\ell}^\ell \left|a_{\ell m}\right|^2.
\end{equation}
In the same way as the multipole coefficients, any significant anisotropy of the angular distribution over scales 
near $1/\ell$ radians would be captured in a non-zero power in the mode $\ell$. Although the exhaustive information 
of the distribution of arrival directions is encoded in the full set of multipole coefficients, the characterisation of any 
important overall property of the anisotropy is hard to handle in a summary plot from this set of coefficients. Conversely, 
the angular power spectrum does provide such a summary plot. In addition, it is possible that for some fixed mode
numbers $\ell$, all individual $a_{\ell m}$ coefficients do not stand above the background noise but meanwhile do
so once summed quadratically.

\begin{figure}[!t]
  \centering					 
 \includegraphics[width=0.9\textwidth]{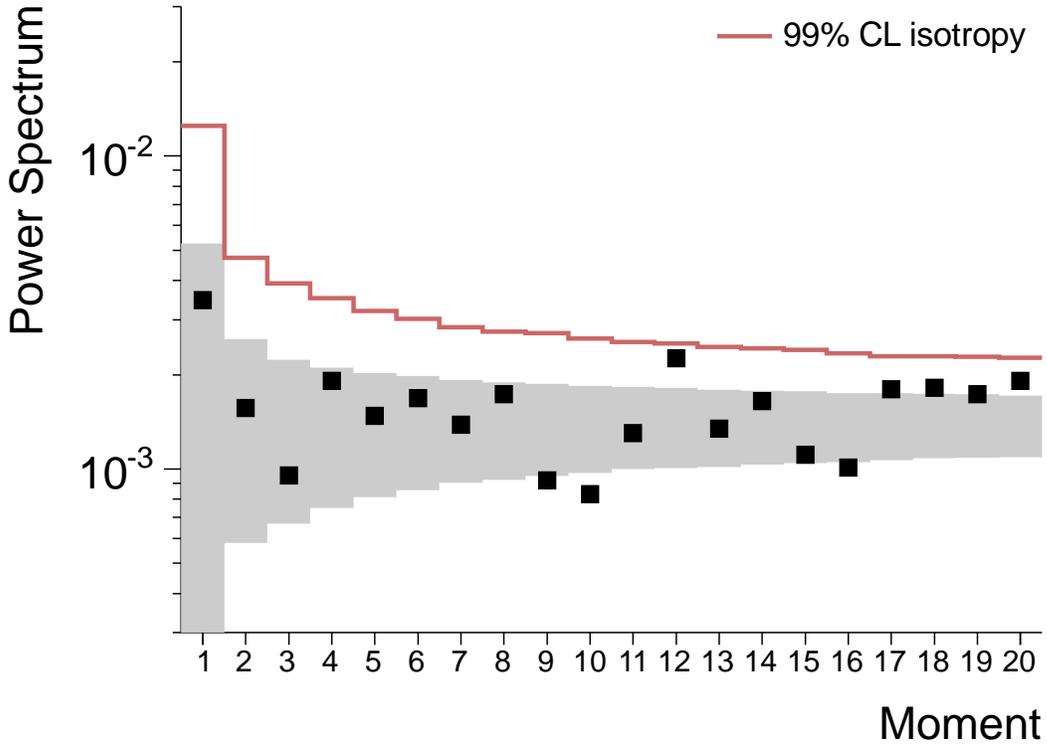}
  \caption{Angular power spectrum.}
  \label{fig:cell}
\end{figure}

From the set of estimated coefficients $\bar{a}_{\ell m}$, the measured power spectrum is shown in figure~\ref{fig:cell}. 
The gray band stands for the RMS of power around the mean values expected from an isotropic distribution, 
while the solid line stands for the 99\% confidence level upper bounds that would result from fluctuations of
an isotropic distribution. Overall, no significant deviation from isotropy is found from this study.

\section{Cross-Checks against Systematic Effects}
\label{sec:systematics}

There are uncertainties in choosing the energy scales to be used when building the joint data set, and/or in
correcting the energy estimator for instrumental effects, and these propagate into systematic uncertainties
in the measured anisotropy parameters. In this section, we choose to use the angular power spectrum as a
relevant proxy to probe the size of the systematic effects investigated below. 

\begin{figure}[!t]
  \centering					 
 \includegraphics[width=0.8\textwidth]{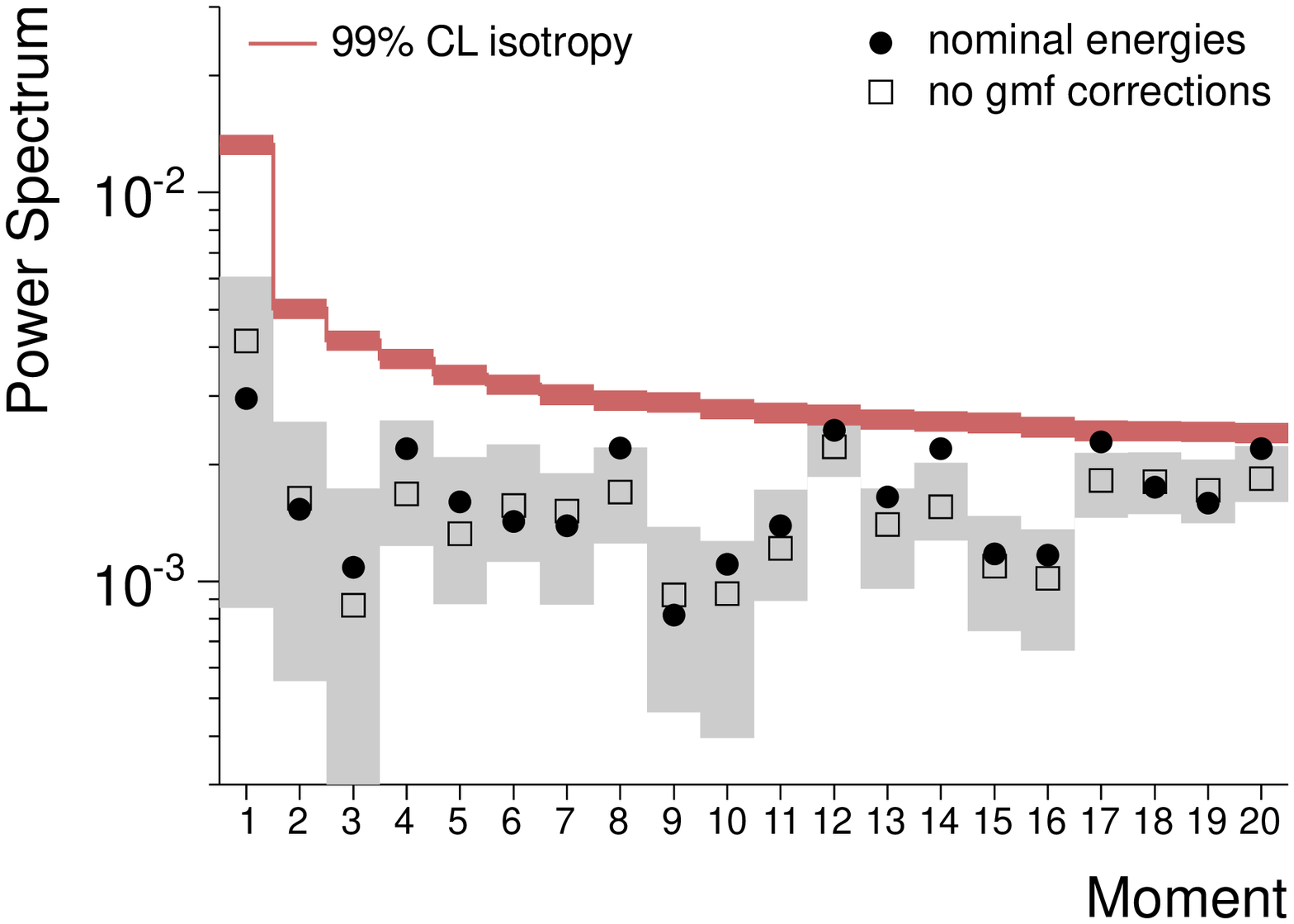}
  \caption{Angular power spectrum as obtained with nominal energies (filled circles) or uncorrected energies
  for geomagnetic effects in the Auger data set (open squares). The values reported in figure~\ref{fig:cell}
  and their statistical uncertainties are indicated by the gray bands.}
  \label{fig:systcell}
\end{figure}

Even though the cross-calibration of energies can be considered as a reasonable starting point for
building the joint data set, it is not a necessary input for the iterative procedure described in 
section~\ref{sec:jointmethod}. Using instead the nominal energies of each experiment only results in a
different balance between the two nominal exposures obtained by requiring equal fluxes in the common 
band through equation~\eqref{eqn:b0}. Then, the final anisotropy parameters necessarily differ, but only slightly.
This is evidenced in figure~\ref{fig:systcell}, where the points (and their statistical uncertainties) of the power 
spectrum obtained previously are shown as the gray bands for each moment $\ell$. The power spectrum 
obtained using nominal energies is shown as the filled circles. The picture is in global agreement, with
only few points standing at most one standard deviation away. 

As already mentioned in section~\ref{sec:rate}, some distortions are imprinted in the event rate as a function of 
local angles by the influence of the geomagnetic field on the development of the showers. The importance of this 
effect depends on the weight of the muonic component to the signal size. At the Pierre Auger Observatory, this 
effect has been shown to induce significant variations of the event rate in declination as soon as the total number 
of events analyzed is of the order of 30\,000~\citep{auger-gmf} if the corrections of the energy estimator discussed 
in section~\ref{sec:rate} are not applied. Given the current statistics above $10^{19}$\,eV, the distortions are
however expected to be marginal for the specific analysis reported here. This is evidenced by the power spectrum
shown as the open squares in figure~\ref{fig:systcell} obtained \emph{without} applying the corrections to the 
(cross-calibrated) Auger data set. All points are indeed observed to be within the statistical uncertainties of the 
estimate shown in figure~\ref{fig:cell}. Given the smaller statistics available in the data set of the Telescope 
Array, and given that the size of this geomagnetic effect is expected to be smaller due to the lower weight of the
muonic component to the signal size with scintillators compared to water-Cherenkov detectors, this provides 
support that the absence of energy corrections in the data set of the Telescope Array does not impact on the results
presented in this report. 

Note that the spread of the 99\% confidence level line in figure~\ref{fig:systcell} stands for the slightly different
statistics which result when using nominal or cross-calibrated energies to select all events above $10^{19}$\,eV.

\section{Conclusions}
\label{sec:conclusions}

In this work, an entire mapping of the celestial sphere has been presented by combining data sets recorded 
at the Telescope Array and the Pierre Auger Observatory above $10^{19}$\,eV. The unavoidable systematic 
uncertainty in the relative exposures has been treated by designing a cross-calibration procedure relying on the 
common region of the sky covered by both experiments. This cross-calibration procedure makes it possible to 
use the powerful multipolar analysis method for characterising the sky map of ultra-high energy cosmic rays. 
Throughout the series of anisotropy searches performed, no significant deviation from isotropy could be 
captured at any angular scale. 

\begin{figure}[!t]
  \centering					 
 \includegraphics[width=0.48\textwidth]{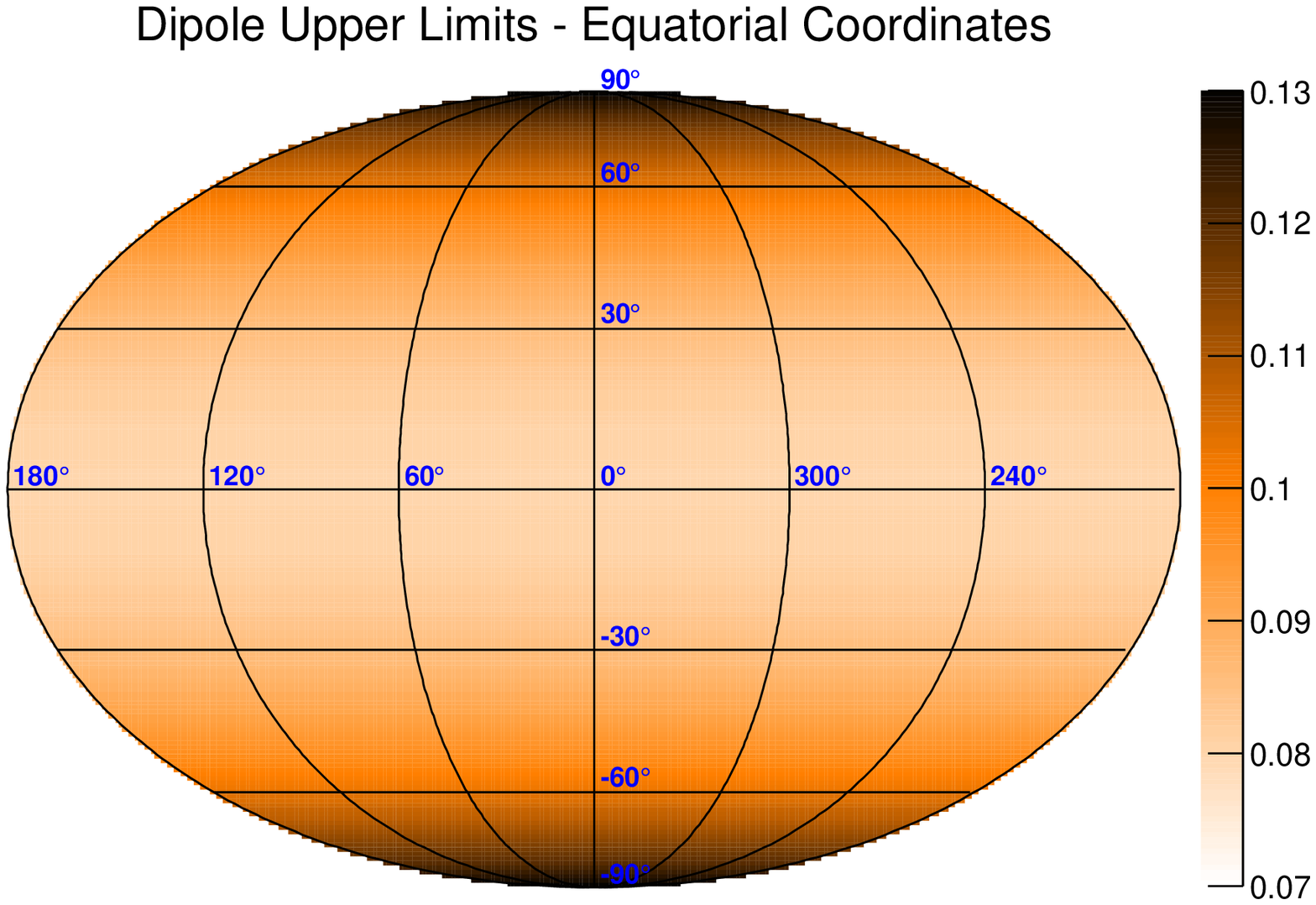}
 \includegraphics[width=0.48\textwidth]{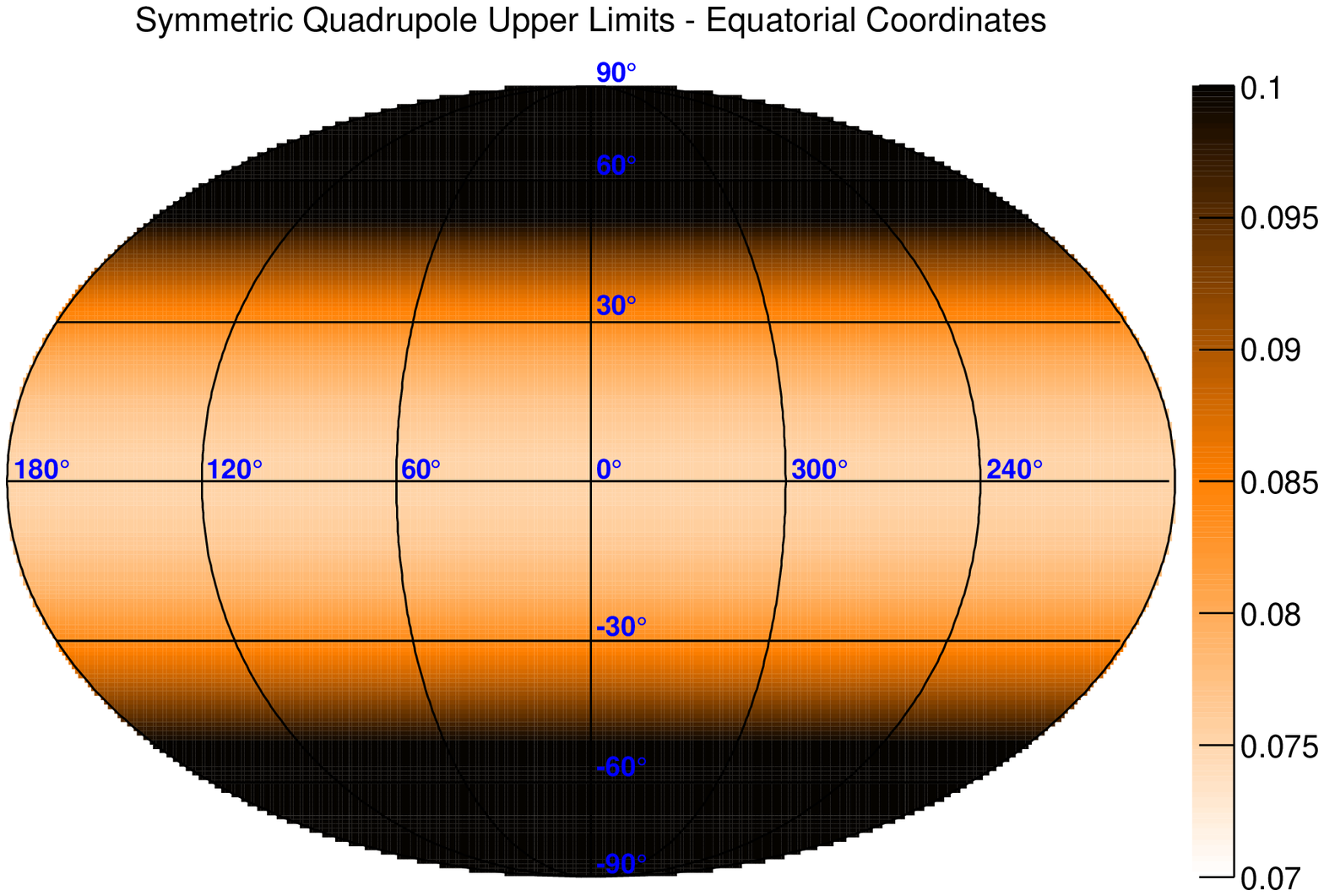}
  \caption{Left: 99\% confidence level upper limits on the dipole amplitude as a function of the latitude
  and longitude, in Equatorial coordinates and Mollweide projection. Right: Same for the amplitude of a symmetric 
  quadrupole.}
  \label{fig:upplim}
\end{figure}

From the multipolar coefficient measurements performed in this work, upper limits can be derived for any kind
of pattern. The ones obtained at 99\% confidence level on the dipole and quadrupole amplitudes are shown in
figure~\ref{fig:upplim} as a function of the direction in the sky, in Equatorial coordinates. These upper limits have 
been obtained by searching for the smallest values of dipole amplitude oriented in each direction $\mathbf{d}$
and quadrupole amplitude oriented in each direction $\mathbf{q}_+$ guaranteeing that the reconstructed 
amplitudes in simulated data sets are larger or equal to the ones obtained for real data in 99\% of the simulations.
The different sensitivities for each direction are caused by the different resolutions for each reconstructed multipolar 
coefficient. Note that the upper limits on the quadrupole amplitude pertain to a symmetric quadrupole only (that is, 
a quadrupole with amplitudes such that $\lambda_-=\lambda_0=-\lambda_+/2$) to keep the number of studied 
variables manageable.

For the first time, the upper limits on the dipole amplitude reported in figure~\ref{fig:upplim} do not rely on 
any assumption on the underlying flux of cosmic rays thanks to the full-sky coverage achieved in this joint study.
With partial-sky coverage, similar sensitivity could be obtained in this energy range but assuming a pure dipolar 
flux~\citep{auger-ls3d}. In addition, the sensitivity on the quadrupole amplitude (and to higher order multipoles 
as well) is the best ever obtained thanks, also, to the full-sky coverage.

The cross-calibration procedure designed in this study pertains to any combined data sets from different 
observatories showing an overlap in their respective directional exposure functions and covering the whole
sky once combined. It is conceivable to apply it in an energy range where the detection efficiency is not 
saturated. Then, future joint studies between the Telescope Array and the Pierre Auger collaborations will allow
further characterisation of the arrival direction distributions down to ${\simeq}10^{18}$\,eV.

\section*{Appendix}
We provide in this appendix the transformation rules between the multipole coefficients and the parameters 
of the dipole vector and the quadrupole tensor. The multipole coefficients are assumed to be calculated from
arrival directions expressed in equatorial coordinates. The Cartesian components of the dipole vector $\mathbf{d}$ 
are related to the $a_{1m}$ coefficients through
\begin{equation}
d_x=\frac{\sqrt{3}}{a_{00}}\,a_{11},\quad d_y=\frac{\sqrt{3}}{a_{00}}\,a_{1-1},\quad d_z=\frac{\sqrt{3}}{a_{00}}\,a_{10}.
\end{equation}
The amplitude $d$ and directions $\delta_d$ and $\alpha_d$ are then obtained by
\begin{equation}
d=\sqrt{d_x^2+d_y^2+d_z^2},\quad \delta_d=\arcsin{d_z},\quad \alpha_d=\arctan{d_y/d_x}.
\end{equation}
The quadrupole can be described by a second order tensor $\mathbf{Q}$ such that the flux can be 
expressed as
\begin{equation}
\Phi(\mathbf{n})=\frac{\Phi_0}{4\pi}\left(1+r\,\mathbf{d}\cdot\mathbf{n}+\tfrac{1}{2}\textstyle{\sum_{i,j}Q_{ij}\,n_i\,n_j}\right).
\end{equation}
$\mathbf{Q}$ is a traceless and symmetric tensor. Its five independent components are determined from the 
$a_{2m}$ by
\begin{align}
Q_{xx} &=  \frac{\sqrt{5}}{a_{00}}\,(\sqrt{3}a_{22}-a_{20}),\\
Q_{xy} &=  \frac{\sqrt{15}}{a_{00}}\,a_{2-2},\\
Q_{xz} &= -\frac{\sqrt{15}}{a_{00}}\,a_{21},\\
Q_{yy} &=  \frac{\sqrt{5}}{a_{00}}\,(\sqrt{3}a_{22}+a_{20}),\\
Q_{yz} &= -\frac{\sqrt{15}}{a_{00}}\,a_{2-1}.
\end{align}
The other components are obtained by symmetry and from the traceless condition (that is, $Q_{zz}=-Q_{xx}-Q_{yy}$).
The amplitudes $\lambda_{\pm,0}$ are then obtained as the eigenvalues of $\mathbf{Q}$ and the vectors $\mathbf{q}_{\pm,0}$
as the corresponding eigenvectors.

\section*{Acknowledgments}

The successful installation, commissioning, and operation of the Pierre Auger Observatory would not have been possible without the strong commitment and effort from the technical and administrative staff in Malarg\"{u}e. We are very grateful to the following agencies and organizations for financial support: 
Comisi\'{o}n Nacional de Energ\'{\i}a At\'{o}mica, Fundaci\'{o}n Antorchas, Gobierno De La Provincia de Mendoza, Municipalidad de Malarg\"{u}e, NDM Holdings and Valle Las Le\~{n}as, in gratitude for their continuing cooperation over land access, Argentina; the Australian Research Council; Conselho Nacional de Desenvolvimento Cient\'{\i}fico e Tecnol\'{o}gico (CNPq), Financiadora de Estudos e Projetos (FINEP), Funda\c{c}\~{a}o de Amparo \`{a} Pesquisa do Estado de Rio de Janeiro (FAPERJ), S\~{a}o Paulo Research Foundation (FAPESP) Grants \# 2010/07359-6, \# 1999/05404-3, Minist\'{e}rio de Ci\^{e}ncia e Tecnologia (MCT), Brazil; MSMT-CR LG13007, 7AMB14AR005, CZ.1.05/2.1.00/03.0058 and the Czech Science Foundation grant 14-17501S, Czech Republic;  Centre de Calcul IN2P3/CNRS, Centre National de la Recherche Scientifique (CNRS), Conseil R\'{e}gional Ile-de-France, D\'{e}partement Physique Nucl\'{e}aire et Corpusculaire (PNC-IN2P3/CNRS), D\'{e}partement Sciences de l'Univers (SDU-INSU/CNRS), Institut Lagrange de Paris, ILP LABEX ANR-10-LABX-63, within the Investissements d'Avenir Programme  ANR-11-IDEX-0004-02, France; Bundesministerium f\"{u}r Bildung und Forschung (BMBF), Deutsche Forschungsgemeinschaft (DFG), Finanzministerium Baden-W\"{u}rttemberg, Helmholtz-Gemeinschaft Deutscher Forschungszentren (HGF), Ministerium f\"{u}r Wissenschaft und Forschung, Nordrhein Westfalen, Ministerium f\"{u}r Wissenschaft, Forschung und Kunst, Baden-W\"{u}rttemberg, Germany; Istituto Nazionale di Fisica Nucleare (INFN), Ministero dell'Istruzione, dell'Univer\\
sit\`{a} e della Ricerca (MIUR), Gran Sasso Center for Astroparticle Physics (CFA), CETEMPS Center of Excellence, Italy; Consejo Nacional de Ciencia y Tecnolog\'{\i}a (CONACYT), Mexico; Ministerie van Onderwijs, Cultuur en Wetenschap, Nederlandse Organisatie voor Wetenschappelijk Onderzoek (NWO), Stichting voor Fundamenteel Onderzoek der Materie (FOM), Netherlands; National Centre for Research and Development, Grant Nos.ERA-NET-ASPERA/01/11 and ERA-NET-ASPERA/02/11, National Science Centre, Grant Nos. 2013/08/M/ST9/00322, 2013/08/M/ST9/00728 and HARMONIA 5 - 2013/10/M/ST9/00062, Poland; Portuguese national funds and FEDER funds within COMPETE - Programa Operacional Factores de Competitividade through Funda\c{c}\~{a}o para a Ci\^{e}ncia e a Tecnologia, Portugal; Romanian Authority for Scientific Research ANCS, CNDI-UEFISCDI partnership projects nr.20/2012 and nr.194/2012, project nr.1/ASPERA2/2012 ERA-NET, PN-II-RU-PD-2011-3-0145-17, and PN-II-RU-PD-2011-3-0062, the Minister of National  Education, Programme for research - Space Technology and Advanced Research - STAR, project number 83/2013, Romania; Slovenian Research Agency, Slovenia; Comunidad de Madrid, FEDER funds, Ministerio de Educaci\'{o}n y Ciencia, Xunta de Galicia, European Community 7th Framework Program, Grant No. FP7-PEOPLE-2012-IEF-328826, Spain; Science and Technology Facilities Council, United Kingdom; Department of Energy, Contract No. DE-AC02-07CH11359, DE-FR02-04ER41300, and DE-FG02-99ER41107, National Science Foundation, Grant No. 0450696, The Grainger Foundation, USA; NAFOSTED, Vietnam; Marie Curie-IRSES/EPLANET, European Particle Physics Latin American Network, European Union 7th Framework Program, Grant No. PIRSES-2009-GA-246806; and UNESCO.

The Telescope Array experiment is supported by the Japan Society for the Promotion of Science through Grants-in-Aids for Scientific Research on Specially Promoted Research (21000002) ``Extreme Phenomena in the Universe Explored by Highest Energy Cosmic Rays'' and for Scientific Research (19104006), and the Inter-University Research Program of the Institute for Cosmic Ray Research; by the U.S. National Science Foundation awards PHY-0307098, PHY-0601915, PHY-0649681, PHY-0703893, PHY-0758342, PHY-0848320, PHY-1069280, and PHY-1069286; by the National Research Foundation of Korea (2007-0093860, R32-10130, 2012R1A1A2008381, 2013004883); by the Russian Academy of Sciences, RFBR grants 11-02-01528a and 13-02-01311a (INR), IISN project No. 4.4509.10 and Belgian Science Policy under IUAP VII/37 (ULB). The foundations of Dr. Ezekiel R. and Edna Wattis Dumke, Willard L. Eccles and the George S. and Dolores Dore Eccles all helped with generous donations. The State of Utah supported the project through its Economic Development Board, and the University of Utah through the Office of the Vice President for Research. The experimental site became available through the cooperation of the Utah School and Institutional Trust Lands Administration (SITLA), U.S. Bureau of Land Management, and the U.S. Air Force. We also wish to thank the people and the officials of Millard County, Utah for their steadfast and warm support. We gratefully acknowledge the contributions from the technical staffs of our home institutions. An allocation of computer time from the Center for High Performance Computing at the University of Utah is gratefully acknowledged.


\end{document}